\documentclass[12pt,twoside]{article}   
\usepackage[textwidth = 18cm]{geometry}
\usepackage{caption}
\usepackage{blindtext}
\usepackage{multirow}
\usepackage{subfiles}

\usepackage[super,square,sort&compress,comma]{natbib}

\usepackage{ragged2e}

\usepackage{fancyhdr}		

\usepackage[export]{adjustbox}
\usepackage{mwe}
\usepackage{framed}
\usepackage{mdframed}
\usepackage{caption}
\usepackage[strict]{changepage}

\usepackage{graphicx}
\usepackage{framed}
\usepackage{tabularx}
\usepackage{array}
\usepackage{makecell}
\usepackage{caption}
\usepackage{floatrow}
\DeclareFloatFont{tiny}{\tiny}

\usepackage{etoolbox}
\makeatletter

\usepackage{sectsty}

\usepackage{etoolbox}
\usepackage{tabularray}
\usepackage{booktabs}

\newcommand{\captionv}[3]{\begin{center}\parbox{#1cm}{\caption[#2]{{\sf #3}}}
        \end{center}}

\usepackage[section]{placeins}   %

\usepackage{graphicx}

\makeatletter \renewcommand\@biblabel[1]{$^{#1}$} \makeatother
\newcommand{\cen}[1]{\begin{center} #1 \end{center}}

\usepackage{caption} 

\usepackage[mathlines,edtable]{lineno}

\usepackage{hyperref}
\hypersetup{ colorlinks,
    citecolor=blue,
    filecolor=blue,
    linkcolor=blue,
    urlcolor=blue
}
\usepackage{titlesec}
\titlespacing*{\subsection}{1pt}{1\baselineskip}{0.1\baselineskip}
\mdfdefinestyle{boxcaption}{innerleftmargin=0.45cm,innerrightmargin=0.45cm}

\titlespacing*{\subsubsection}{1pt}{1\baselineskip}{0.1\baselineskip}

\usepackage{xcolor}

\definecolor{gray}{rgb}{0.6,0.6,0.6}
\definecolor{red}{rgb}{0.85,0,0}
\definecolor{green}{rgb}{0,0.85,0}
\definecolor{blue}{rgb}{0,0,0.85}
\definecolor{beige}{rgb}{0.92,0.87,0.78}
\usepackage[all]{hypcap}    
\usepackage{setspace}

\begin{document}

\cen{\sf {\LARGE {\bfseries  Quantification of head and neck cancer patients' anatomical changes during radiotherapy: prediction of replanning need} 
\vspace{5mm}

\small{Odette Rios-Ibacache$^{1}$,
James Manalad$^{1}$,
Kayla O'Sullivan-Steben$^{1}$,
Emily Poon$^{2}$,
Luc Galarneau$^{2}$,
Julia Khriguian$^{3}$,
George Shenouda$^{4}$ \&
John Kildea$^{1,2}$}

$^{1}$Medical Physics Unit, McGill University, Montreal, QC, Canada\\
$^{2}$Research Institute of the McGill University Health Centre, McGill University, Montreal, QC, Canada\\
$^{3}$Department of Radiation Oncology, Hôpital Maisonneuve-Rosemont, Montreal, QC, Canada. \\
$^{4}$Department of Radiation Oncology, McGill University, Montreal, QC, Canada.
\vspace{5mm}\\
Version typeset \today\\}}

\pagenumbering{roman}
\setcounter{page}{1}
\pagestyle{plain}
\noindent \textbf{keywords:}\textit{treatment replanning, radiotherapy, geometrical metrics, machine learning, head and neck cancer}\\
Correspondence author email: odette.riosibacache@mail.mcgill.ca \\

\vspace{-3mm}
\begin{abstract}
\noindent {\bf Background:} Head and neck cancer (HNC) patients who undergo radiotherapy (RT) may experience anatomical changes during treatment, which can compromise the validity of the initial treatment plan, necessitating replanning. However, ad hoc replanning disrupts clinical workflows, creating a stressful environment. Currently, no standardized method exists to determine the total amount of anatomical variation that necessitates replanning.\\ 
{\bf Purpose:} This project aimed to create metrics to describe anatomical changes HNC patients may experience during RT. The usefulness of these metrics was evaluated by developing machine learning (ML) models to predict the need for replanning.\\
{\bf Methods:} This study included a cohort of 150 HNC patients treated at the McGill University Health Centre. Based on the shape of the RT structures, we created 43 metrics and developed an extraction pipeline in Python, called HNGeoNatomyX, to automatically calculate them. A univariate metric analysis using linear regression was conducted to obtain the rate of change of each metric. We also obtained the relative variation of each metric between the pre-treatment scan and the fraction at which replanning was requested. Fraction-specific ML models (models that incorporated information available up to and including the specific fraction) for fractions 5, 10, and 15 were built using the metrics, clinical data, and feature selection techniques. To estimate the performance of the models, we used a repeated stratified 5-fold cross-validation resampling technique and the Area Under the Curve (AUC) of the Receiver Operating Characteristic (ROC) curve.  \\
{\bf Results:} The best specific multivariate models for fractions 5, 10, and 15 yielded testing scores of 0.82, 0.70, and 0.79, respectively. Our models early predicted replanning for 76\% of the true positives.  \\
{\bf Conclusions:} The created metrics have the potential to characterize and distinguish which patients will necessitate RT replanning. They show promise in guiding clinicians to evaluate RT replanning for HNC patients and streamline workflows. \\

\end{abstract}



\setlength{\baselineskip}{0.6cm}      
\pagenumbering{arabic}
\setcounter{page}{1}
\pagestyle{fancy}

\setstretch{1}
\section{Introduction}
\vspace{-3mm}


Worldwide, more than 830,000 people are diagnosed with Head and Neck Cancer (HNC), and more than 430,000 patients die from it each year \cite{Amaral2022-eo}. Statistically, around 75\% of patients with HNC have been reported to benefit from RT treatment as a primary or adjuvant treatment option \cite{Gregoire}. However, patients with HNC who undergo RT may experience acute or late radiation toxicities during or after treatment, respectively, including dermatitis, xerostomia, and mucositis \cite{Tsai2023-fb, Larsson2003-bh}. These toxicities may be painful and contribute to local infections. They can also affect nutrition, as patients may have difficulty eating, drinking, and swallowing \cite{Sroussi2017-fz}, which can result in weight loss and anatomical changes (volumetric or spatial changes). Furthermore, the tumor itself may respond well to radiation and shrink in size. 

Anatomical changes during HNC RT are problematic, as they can compromise the delivery of the prescribed treatment dose (i.e., underdosing the tumor volume or overdosing the organs at risk (OARs)), necessitating treatment replanning \cite{Barker2004-pl, Jham2006-ep}. For example, due to the weight loss that patients may experience, the Planning Target Volume (PTV) contour considered initially can be invalidated. Additionally, the thermoplastic treatment mask used for immobilization can become loose. The mask looseness can, in turn, lead to unwanted movement and positioning difficulties, causing setup errors during treatment, which can then affect the delivered doses to many critical structures, such as the brainstem and spinal cord \cite{Robar2007-uc}.

In our clinic, the McGill University Health Centre (MUHC) in Montreal, Canada, approximately 9\% of HNC patients treated with RT require replanning at some point during RT treatment each year. According to the literature \cite{Figen2020-qz, Chinnery2024-pz, Brown2015-ut, Van_Beek2019-hc}, 3\% - 10.6\% of HNC patients require a replan. Figen et al. (2020) \cite{Figen2020-qz} reported that the most common replanning requests in the clinic are due to tumor shrinkage and weight loss, accounting for 71\% of the requests. 

Despite the widespread need to replan, there is no formal consensus or standardized guidance on when to replan in HNC RT. For example, Bhide et al. (2010) \cite{Bhide2010-vh} noted that the parotid glands received an increased dose by the 4th week of treatment due to their medial shift, indicating that patients could benefit from replanning before week four. In contrast, Fiorentino A. et al. (2012)\cite{Fiorentino2012-kd} suggested that a replan should be indicated by the 3rd week of treatment to avoid parotid gland overdose. Wang et al. (2010) \cite{Wang2010-al} observed that, for patients with nasopharyngeal HNC, replanning before the 25th fraction of RT ensures safe doses to OARs. 

Furthermore, no standard method has been established to define a threshold amount of mass loss or anatomical change in the head and neck region that triggers a replan \cite{Morgan2020-wc}. The planning workflow of a clinic is typically complex, and various workloads and economic factors\cite{WALLS2023e632} can cause the replanning process to be delayed until absolutely necessary \cite{Heukelom2019-dm}. Clinical personnel, including radiation therapists and medical physicists, must work together to schedule and perform CT rescanning, recontouring, and replanning effectively \cite{Volpe2021-cq}.

To address this issue, some studies, such as Brown et al. (2016)\cite{Brown2016-ly}, have tried to predict the \textit{optimal time} for replanning by attempting to identify \textit{pre-treatment factors} that influence the replanning decision. Their results showed that nasopharyngeal cases require replanning earlier than oropharyngeal cases (3rd and 4th weeks of treatment, respectively). In the context of the possible use of artificial intelligence (AI) methods to predict RT replanning for HNC, a recent study by Chinnery et al. (2024)\cite{Chinnery2024-pz} showed that radiomic and dosimetric information from initial simulation CT (CT sim) scans has the potential to aid replanning predictions. However, their model did not incorporate information on the effect of RT on the patient and did not consider subsequent medical images after the initiation of treatment. 

The lack of a tool that can systematically record and characterize the anatomical and dosimetric changes that HNC patients experience during RT makes it difficult to assess which patients may benefit from replanning and when. With this in mind, our study aimed to create metrics that quantify anatomical changes in the HN region and examine their predictive values, in combination with clinical data, for RT replanning in HNC patients using a machine learning (ML) model. We surmised that the development of such metrics and the creation of an automatic pipeline to calculate and evaluate them could serve as a supportive decision aid for clinicians during the treatment of HNC patients, incorporating the effect of RT.
\vspace{-3mm}

\section{Materials and methods}
\vspace{-5mm}

\subsection{Patient cohort and study sample}

This retrospective study considered patients diagnosed with HNC or unknown primary tumors in the head and neck region. Data were collected for a cohort of 381 patients who received or started receiving RT treatment at the MUHC between January 1, 2017 and March 31, 2024. Treatment regimens included RT alone or in combination with systemic therapies, such as chemotherapy and targeted therapy. The prescribed RT doses in our patient cohort consisted of 70 Gy in 35 fractions, 60 Gy in 30 fractions, and 66 Gy in 33 fractions to the PTV, with curative intent. This retrospective study was approved by the Research Ethics Board (REB) of the McGill University Health Centre, Montreal, Quebec, Canada. All work of the study was conducted in accordance with REB guidelines.

Patients' inclusion and exclusion criteria were established based on the number of CBCT images acquired, the reported reasons for replanning, and whether or not patients finished their treatments. Since our institutional imaging guidelines specify that two CBCTs should be taken per week and considering that anatomical changes have been reported to occur during the first week of treatment \cite{Delaby2023-or}, each patient was required to have a CT sim image and at least two CBCT images taken on two different fractions during treatment. The replan request fraction had to be within the fraction range of the first and last CBCT images or correspond to one of them. Furthermore, we excluded cases where the reported replanning reasons were not related to the patient's anatomy. These reasons include holes that developed in the treatment masks, patients who did not tolerate the treatment mask, or patients who were unable to tolerate treatment for psychological reasons. The dataset selection process is summarized in Figure A of the Supplementary Material. 

Among the 362 eligible patients, 37.8\% (n=137) had their RT treatments replanned, and 62.2\% (n=225) completed their RT course without replanning. Due to time constraints, a sample of 75 replanned cases and 75 non-replanned cases was randomly selected from the eligible patient cohort to provide a balanced class distribution for the ML analysis\cite{Oommen2011-jo}. 

\subsection{Data collection}

CT sim images, taken at the beginning of treatment planning, and CBCT images, taken over the course of treatment, were collected from the Treatment Planning System (TPS), Eclipse v.15 (Varian Medical Systems, Inc., Palo Alto, CA). All medical images were exported from the TPS as DICOM files, and all RT data were exported as DICOM-RT files. All data were anonymized during the TPS export.

\begin{figure*}[h]
\vspace{-5mm}
\captionv{12}{Short title - can be blank}{
\justifying{Figure A shows the workflow describing the steps that were followed for the metrics definition and statistical analysis, beginning with the image registration process and finishing with the machine learning analysis. Figure B shows the generated 3D structures used for our analysis, including the CT sim and CBCTs body contouring, and geometrical metrics definition.
\label{Wokflow}}}
\includegraphics[width=1\linewidth]{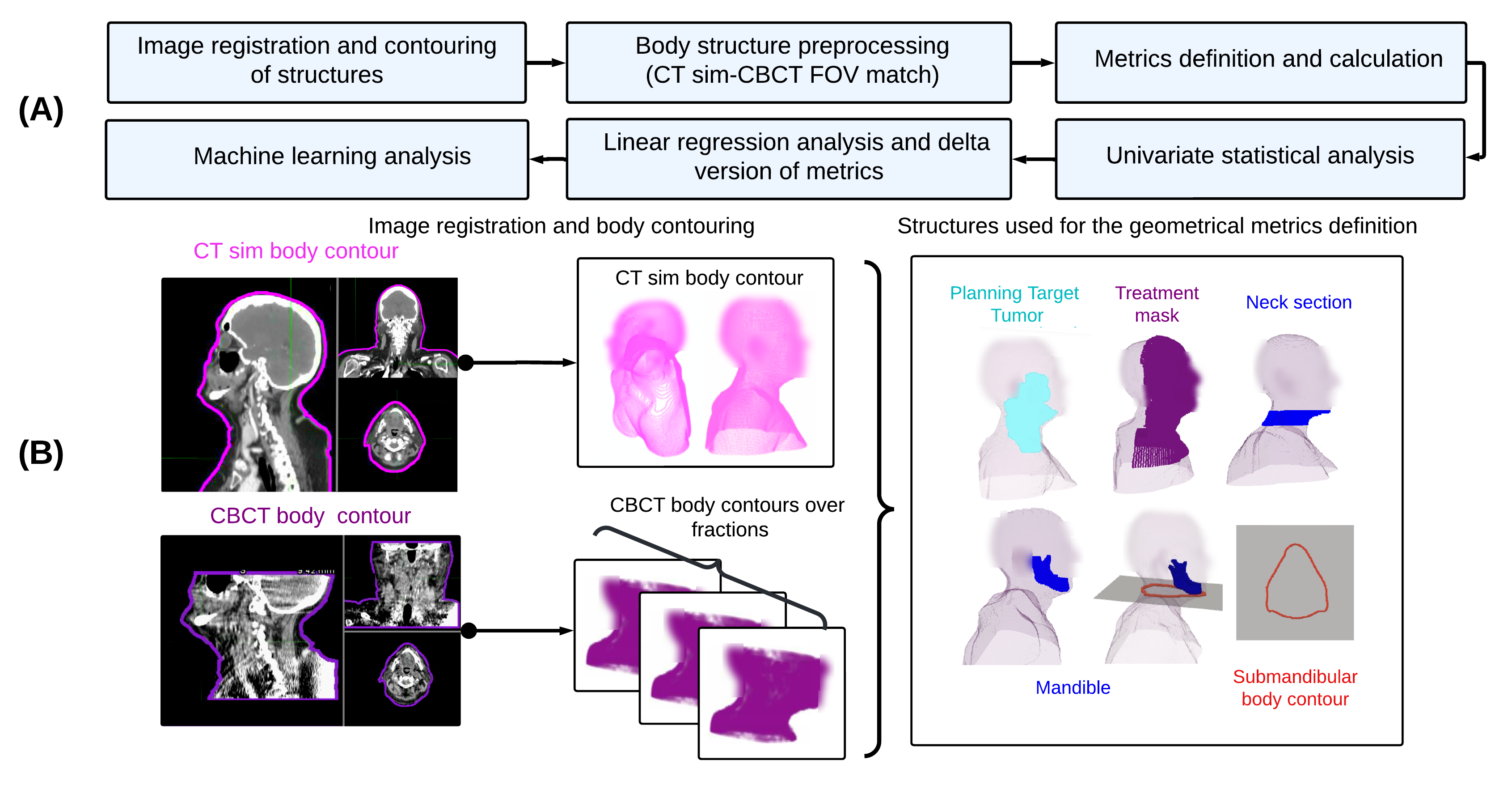}
\end{figure*}

\vspace{2mm}
In our cancer centre, clinicians use questionnaires, recorded at regular intervals throughout RT treatment, to collect certain temporally variable clinical information for HNC patients. These data include weight, toxicity grades (e.g., dysphagia, mucositis, xerostomia, dermatitis, and laryngeal side effects), Karnofsky Performance Status (KPS), hematological test results, and presence or absence of tube feeding (percutaneous endoscopic gastrostomy (PEG)). These data were included to allow a thorough evaluation of the clinical evolution of each patient over time.

This study also incorporated time-independent clinical information about the patient, including the status of the tumor suppressor gene p16 (a surrogate marker for HPV status \cite{Da_Mata2021-wk}), smoking or tobacco use history, TNM staging, overall stage of cancer, and age at the start of treatment.

To study the anatomical changes that occurred during RT treatment, geometrical information was analyzed using medical images, RT structures, and patient body shapes. The following subsections describe the process of contouring the external body and the treatment mask, which were then used to create metrics to characterize anatomical changes. Figure \ref{Wokflow} summarizes the workflow of this study.

\vspace{-3mm}
\subsection{Image registration and contouring of structures}

For each patient, their CBCT DICOM images were registered to their CT sim DICOM image to align with the RT structures corresponding to the PTV and OARs, using a rigid registration method. For the replanned patients, only their pre-replanning CBCTs were registered. The body contour of each CBCT image was generated using the commercial contouring application MIM MAESTRO v7.1.6 (MIM Software Inc., Cleveland, OH). To ensure that the body structure was correctly contoured, each slice was visually examined. Slices in which the patient's imaged anatomy was partially imaged were not included in the body contour. The final contoured image was saved in the DICOM-RT structure set format. Since the goal of this study was to predict replanning and considering that replanning often occurs before the 4th week of treatment\cite{Bhide2010-vh,Zhao2011-ek}, the registration process was only performed for fractions before or including the 25th fraction.

Although treatment mask looseness is sometimes the reason for replanning\cite{Figen2020-qz}, the mask is not included as an RT structure in clinical practice. To address this, a semiautomatic 2D contouring algorithm was developed in Python v3.7 to define the mask structure. The algorithm receives as input the array of the CT sim image slices, the index number of the slice to be contoured, the CT sim body contour (used to calculate the superior and inferior limits), and a threshold value. The output points are saved in the same coordinate system as the RT structure set of each patient. The resulting matrix of points is stored in the \textit{JSON} file format. The summarized steps can be seen in Figure \ref{FOV_reconstruction}(A).

\vspace{-3mm}
\subsubsection{Body structure preprocessing}

Since the cone-shaped beam used to acquire a CBCT image limits its Field Of View (FOV), portions of a patient's anatomy may be missing from the CBCT when compared to the corresponding CT sim image, which has a much wider FOV. Thus, to ensure consistent anatomy across fractions for a given patient, a preprocessing step was applied to reconstruct the CBCT FOV and use it to crop the CT sim's contour.

\begin{figure}[h]
\captionv{12}{Short title - can be blank}{
\justifying{(A) Flowchart of the treatment mask contouring steps. (B) A diagram describing the CBCT image FOV reconstruction procedure and CT sim contour cropping procedure to match the geometrical shape.\vspace{5mm}}
\label{FOV_reconstruction}}
\centering
\includegraphics[width=0.95\linewidth]{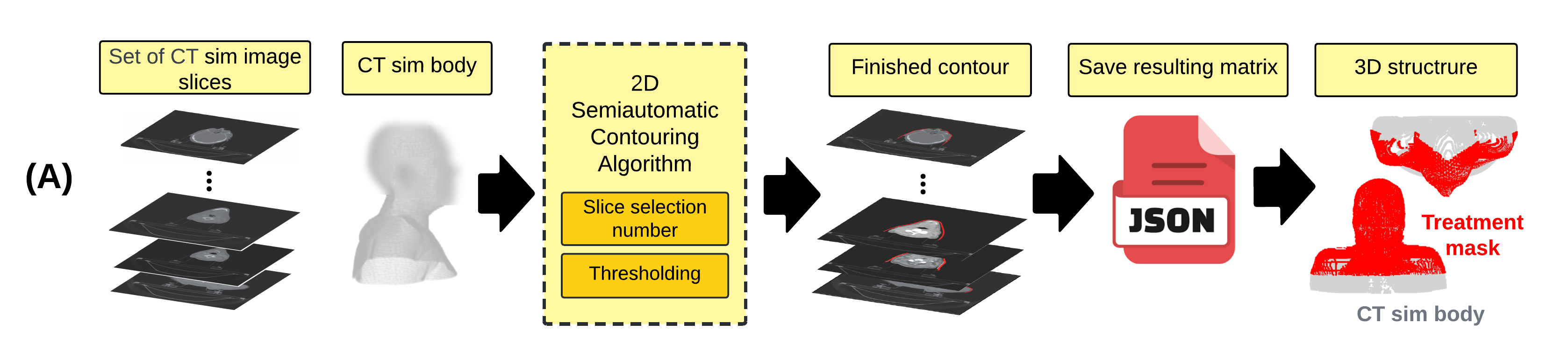}
\includegraphics[width=1.\linewidth]{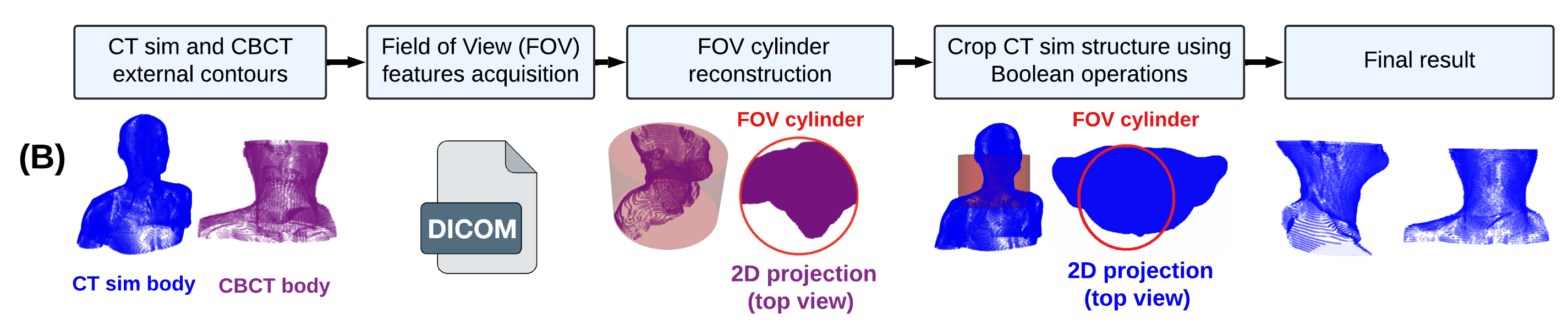}
\end{figure}

The acquisition isocenter of the first CBCT (fraction 1) was extracted from its DICOM-RT Structure Set file (as previously registered to the CT sim using MIM MAESTRO). The reconstruction diameter stored in the CBCT DICOM information was retrieved using the open-source library \textit{pydicom}\cite{Mason2023-pe} to define the outer boundary and reconstruct the FOV. Next, an algorithm based on Boolean operations was created to crop the CT sim's body contour for each patient. The process can be seen in Figure \ref{FOV_reconstruction} (B). Additionally, since each CBCT image captures the anatomy for slightly different z-ranges (height), each body contour was trimmed in the z direction to the common overlapping z-region across all CBCTs for a given patient.

\subsection{Metrics to describe patient anatomy}

Based on the 3D shape and the 2D contours of the body and RT structures (including the PTV and mandible), we defined a total of 43 quantitative and continuous metrics to describe the anatomical changes of the patient during RT delivery. These changes include those observed from CT sim to CBCT, as well as between successive CBCTs. We grouped the metrics into six categories, which are summarized in Table \ref{TableParam1} and detailed in the following subsections. 

\subsubsection{Body-related metrics}

This set of six metrics was based on the body region covered and the differences between each CBCT body contour and the corresponding CT sim body contour. These metrics are labeled using the subscript \textit{Body} in Table \ref{TableParam1}.

We defined a volume metric ($V_{Body}$) to quantify the volume enclosed by the body contour. This metric was calculated on the basis of the total number of voxels enclosed by the body/external contour multiplied by the voxel volume (based on the CT sim image resolution: pixel spacing and slice thickness).

Next, a series of 3D and 2D distances was established to measure the change for each fraction relative to the CT sim. These metrics were inspired by Bivrio D. et al.(2018)\cite{Brivio_2018}, who calculated the differences between body contours in cylindrical coordinates, and the analysis that physicists and radiation oncologists perform during clinical evaluation (a slice-by-slice analysis of images). For the 3D distance $CD_{Body}$, we used the \textit{chamfer distance} module from the open-source library \textit{point-cloud-utils}\cite{point-cloud-utils}. The 3D maximum distance was calculated by obtaining the maximum value of the distances between the nearest neighbours of the points in the contour using \textit{KDtree} from \textit{scipy}\cite{2020SciPy-NMeth}. 

To calculate 2D distances, each axial slice of the contour was analyzed individually. In each slice, the distance between the points was calculated using the \textit{directed hausdorff} module from \textit{scipy}. The maximum, mean, and median distances for each CBCT were determined from the distribution of distances across all slices for that CBCT.

\vspace{5mm}
\begin{table}[h!]
\vspace{-8mm}
\captionv{12}{}{\justifying{Metric definitions that were created to analyze the anatomical changes of replanned and non-replanned patients. \label{TableParam1}}}
\begin{center}
\begin{tabular}{ll|ll}
\specialrule{1.pt}{0.5pt}{0.5pt}
\textbf{Metric} & \textbf{Symbol} & \textbf{Metric} & \textbf{Symbol} \\ \hline
\vspace{0.1mm}
\textit{(1) Body-related metrics}     &   &   \makecell*[l]{Avg. distance from the mandible to  body}& \makecell*[l]{$M_{avg}$} \\
\vspace{-1.5mm}
\makecell*[l]{Body volume}& \makecell*[l]{$V_{Body}$} & \makecell*[l]{SD of distances from the mandible to body}& \makecell*[l]{$\sigma_{M}$}\\
\vspace{-1.5mm}
\makecell*[l]{Chamfer or avg. distance (3D)}& \makecell*[l]{$CD_{Body}$} & \makecell*[l]{}& \makecell*[l]{}\\ \cline{3-4}
\vspace{-1.5mm}
\makecell*[l]{Haussdorff or max. distance (3D)}& \makecell*[l]{$HD_{Body}$} &  \makecell*[l]{\textit{(5) Neck-related metrics}}& \makecell*[l]{} \\ 
\vspace{-1.5mm}
\makecell*[l]{Max. 2D distance}& \makecell*[l]{$D_{Body}$} & \makecell*[l]{Neck volume}& \makecell*[l]{$V_{Neck}$}  \\
\vspace{-1.5mm}
\makecell*[l]{Median 2D distance}& \makecell*[l]{$\tilde{D}_{Body}$} & \makecell*[l]{Chamfer or avg. neck distance}& \makecell*[l]{$CD_{Neck}$}  \\
\vspace{-1.5mm}
\makecell*[l]{Avg. 2D distance}& \makecell*[l]{$\bar{D}_{Body}$} & \makecell*[l]{Haussdorff or max. neck distance}& \makecell*[l]{$HD_{Neck}$}  \\
\cline{1-2}
\textit{(2) Treatment mask-related metrics}  &   & \makecell*[l]{Max. 2D neck distance}& \makecell*[l]{$D_{Neck}$}  \\
\vspace{-1.5mm}
\makecell*[l]{Max. distance to treatment mask}& \makecell*[l]{$max\{B_{mask}\}$} & \makecell*[l]{Median 2D neck distance}& \makecell*[l]{$\tilde{D}_{Neck}$}\\
\vspace{-1.5mm}
\makecell*[l]{Avg. distance to treatment mask}& \makecell*[l]{$\bar{B}_{mask}$} & \makecell*[l]{Avg. 2D neck distance}& \makecell*[l]{$\bar{D}_{Neck}$}\\ 
\vspace{-1.5mm}
\makecell*[l]{SD of distances to treatment mask}& \makecell*[l]{$\sigma_{B_{mask}}$} & \makecell*[l]{Min. 3D neck radius}& \makecell*[l]{$R^{3D}_{min}$}\\
\vspace{-1.5mm}
\makecell*[l]{Air volume between body and \\treatment mask}& \makecell*[l]{$V^{air}_{Body-to-mask}$} & \makecell*[l]{Max. 3D neck radius}& \makecell*[l]{$R^{3D}_{max}$}\\\cline{1-2}
\textit{(3) PTV-related metrics}  &   & {Avg. 3D neck radius}& \makecell*[l]{$R^{3D}_{avg}$}   \\
\vspace{-1mm}
\makecell*[l]{Min. distance from CT sim PTV to \\CBCT body}& \makecell*[l]{$x^{PTV}_{min}$} & 
\makecell*[l]{Ratio between min. and max. 3D \\neck radius}& \makecell*[l]{$\varphi^{3D}_{R}$}\\
\vspace{-1.5mm}
\makecell*[l]{Max. distance from CT sim PTV to \\CBCT body}& \makecell*[l]{$x^{PTV}_{max}$} & 
\makecell*[l]{Avg. cross-sectional neck area}& \makecell*[l]{$A^{2D}_{avg}$}\\
\vspace{-1.5mm}
\makecell*[l]{Avg. distance from PTV to body}& \makecell*[l]{$x^{PTV}_{avg}$} & 
\makecell*[l]{Neck surface area}& \makecell*[l]{$SA_{Neck}$} \\
\vspace{-1.5mm}
\makecell*[l]{Median distance from PTV to body}& 
\makecell*[l]{$x^{PTV}_{med}$} & \makecell*[l]{Neck compactness}& \makecell*[l]{$C_{Neck}$}  \\\cline{3-4}
\vspace{-1.5mm}
\makecell*[l]{SD of the distances from  PTV to body}& 
\makecell*[l]{$x^{PTV}_{std}$} & \makecell*[l]{\textit{(6) Submandibular-related metrics}}& \makecell*[l]{}\\
\makecell*[l]{Volume PTV inner}& \makecell*[l]{$VI_{PTV}$} & 
\makecell*[l]{Submandibular area}& \makecell*[l]{$A_{sub}$} \\
\vspace{-1.5mm}
\makecell*[l]{Volume PTV outer}& \makecell*[l]{$VO_{PTV}$} & \makecell*[l]{Min. 2D submandibular radius}& \makecell*[l]{$R^{sub}_{min}$}\\
\vspace{-1.5mm}
\makecell*[l]{Volume PTV inner ratio}& \makecell*[l]{$VI_{PTV}:LV_{Body}$} & 
\makecell*[l]{Max. 2D submandibular radius}& \makecell*[l]{$R^{sub}_{max}$}\\
\vspace{-1.5mm}
\makecell*[l]{Volume PTV outer ratio}& \makecell*[l]{$VO_{PTV}:LV_{Body}$} & \makecell*[l]{Avg. 2D submandibular radius}& \makecell*[l]{$R^{sub}_{avg}$} \\
\cline{1-2}
\makecell*[l]{\textit{(4) Mandible-related metrics}}& \makecell*[l]{} &\makecell*[l]{Ratio between $R^{sub}_{min}$ and $R^{sub}_{max}$}& \makecell*[l]{$\varphi^{2D}_{R^{sub}}$} \\
\vspace{-1.5mm}
\makecell*[l]{Min. distance from the mandible to body}& \makecell*[l]{$M_{min}$} &  \makecell*[l]{Maximum longitudinal chord}& \makecell*[l]{$l^{sub}_{y}$} \\
\vspace{-1.5mm}
\makecell*[l]{Median distance from the mandible to body}& \makecell*[l]{$M_{med}$}  & \makecell*[l]{Maximum lateral chord}& \makecell*[l]{$l^{sub}_{x}$}\vspace*{1mm}\\
\specialrule{1.pt}{0.5pt}{0.5pt}
\end{tabular}
\end{center}
\end{table}
\vspace{-3mm}

\subsubsection{Treatment mask-related metrics}

A 3D structure of the treatment mask was generated using \textit{pyvista}\cite{sullivan2019pyvista}, based on its contour. Next, to track the looseness of the mask, we created metrics to relate the treatment mask to the body contour. Four metrics were defined for this purpose: three based on distances that were calculated using the \textit{KDTree} and one volume that describes the space between the body and the mask, which was calculated using the same approach as for $V_{Body}$. The metrics were labeled with the subscript \textit{mask}.

\vspace{-5mm}

\subsubsection{PTV-related metrics}

Motivated by Bartosz B. et al. (2022)\cite{Bak2022-bv}, who found that the extension of the PTV outside the body on CBCTs is the reason for a high number of replanning decisions, we defined metrics to describe the location of the PTV (originally generated on the CT sim image) relative to the body. These served as indicators of potential inadequate target coverage during treatment. Five metrics were created to describe the distance between the PTV and the body contour, as well as the portions of the PTV inside and outside the body.

To determine the portion of the PTV inside and outside the body, the \textit{compute implicit distance} and \textit{threshold} modules from the \textit{pyvista} library were used. These functions divide a 3D structure according to a surface, allowing the identification of inner and outer PTV points to calculate the distances. The PTV-related metrics were labeled with the subscript \textit{PTV} in Table \ref{TableParam1}.
\vspace{-3mm}

\subsubsection{Mandible-related metrics}

Another way to characterize the mass loss that patients with HNC may experience is to track the distance between the body contour and a rigid bone structure that experiences minor dimensional changes. This type of bone structure should remain or be fixed in the same position within the body over the course of treatment. The mandible is one of these types of bone structure, and since it was included in the CT sim RT Structure Set file for all patients in our cohort, it served as a reliable reference point to track mass loss in the face region. For each patient, distances were calculated between the mandible and the body contour of the patient. The \textit{compute implicit distance} and \textit{threshold} modules from the \textit{pyvista} library were again used for this purpose. The metrics were labeled with the letter and subscript \textit{M}.
\vspace{-3mm}

\subsubsection{Neck-related metrics}

Inspired by studies that have shown that patients undergoing RT experience changes in the dimensions of their neck regions \cite{Kumar2024-wr, Ahn2011-kg}, our study included the creation of an algorithm to determine consistent boundaries of the neck region, as it was not previously explicitly contoured. The procedure involved the identification of a $z_{b}$ value from the most inferior slice of the body point cloud that does not include any shoulder anatomy and is at the boundary of the FOV cylinder. To consider a consistent neck region across the CBCTs of an individual patient, the region was defined at the superior end by the slice below the mandible and at the inferior end by the most inferior point determined from the CBCT body contours' $z$-values,  $z_{b}$. To avoid the inclusion of the shoulders, a three-slice margin was employed above the most inferior point. Metrics were defined to describe the size and shape of the neck using similar calculation methods used for the body-related metrics. They are presented in Table \ref{TableParam1} using the subscript \textit{Neck}.
\vspace{-3mm}

\subsubsection{Submandibular-related metrics}

To develop comprehensive variables that describe changes in all possible directions and facilitate a thorough analysis, additional 2D metrics related to the submandibular region were created. The submandibular region is part of the neck and can experience a high rate of change during treatment. To calculate these metrics, the submandibular plane was determined along with the corresponding body contour points. To distinguish the metrics, the symbols were labeled with the superscript or subscript \textit{sub}, as appropriate.  

To systematically track changes over time using all defined metrics, an automatic extraction pipeline (referred to as \textit{HNGeoNatomyX}) was developed using the Python v3.7 scripting language. This tool facilitated the management, extraction, and analysis of all the metrics described above. 

\vspace{-5mm}

\subsection{Univariate analysis}

Using our automatic metrics extraction pipeline and a CPU corresponding to an Intel Core Processor (Skylake) with 8 GB of RAM, we analyzed the data of our 150 patients. A univariate statistical analysis was performed to provide information on the potential predictive power of each extracted metric and its ability to guide replanning. This type of analysis included a linear regression and a delta analysis. Linear regression was used to individually examine the evolution of each metric over time, while delta analysis was performed to quantify the anatomical change at the time of the replanning request.

\subsubsection{Linear regression analysis}

This analysis involved the calculation of the rate of change of a given metric to see how it changed across fractions for each of the two classes of patients: replanned and non-replanned. Specifically, the rate of change was obtained from the slope calculation up to each fraction number. As reported in previous studies \cite{El-Sayed2011, Stauch2020-en, Andrade-Hernandez2020-ad}, weight loss and decreased neck area over time were more prominent in replanned patients compared to non-replanned patients. Thus, we hypothesized that the rate at which the metrics change may provide useful information.

We generated graphs to visually inspect the longitudinal data across treatment fractions and show the metrics' progression over time. Quantitatively, each metric was analyzed using the non-parametric Mann-Whitney U (MWU) test to determine if the distributions corresponding to the classes replanned versus non-replanned were statistically different. A Bonferroni correction was applied to account for the application of multiple hypothesis tests \cite{Armstrong2014-zi}. Using the definition of the MWU test, the predictive power of each metric was calculated for the fractions for which the test had \textit{p-values} below 0.05, and an average was obtained. The predictive power was calculated using the AUC value from the $U-$ statistic for ROC curves\cite{articleauc}. A 95\% confidence interval (C.I.) was established through bootstrapping with 100 repetitions. The same procedure was used to compare the results for the weight values (also collected over time).

\begin{figure}[h]
\vspace{-5mm}
\captionv{12}{Short title - can be blank}{
\justifying{{ML pipeline analysis conducted in our research. CV is cross-validation, and GridSearchCV is the hyperparameter tuning method. StandardScaler: Data scaling or standardization method used. LR: Logistic Regression, KNN: KNearestNeighbors, DT: DecisionTree, NB: NaiveBayes, SVM: Support Vector Machine, RF: Random Forest, ET: ExtraTree, XGB: eXtreme Gradient Boost classifier, fxs: fractions, fx$_{c}$ certain fraction of interest.}} 
\vspace{5mm}
\label{ML_flow}}
\centering
\includegraphics[width=1\linewidth]{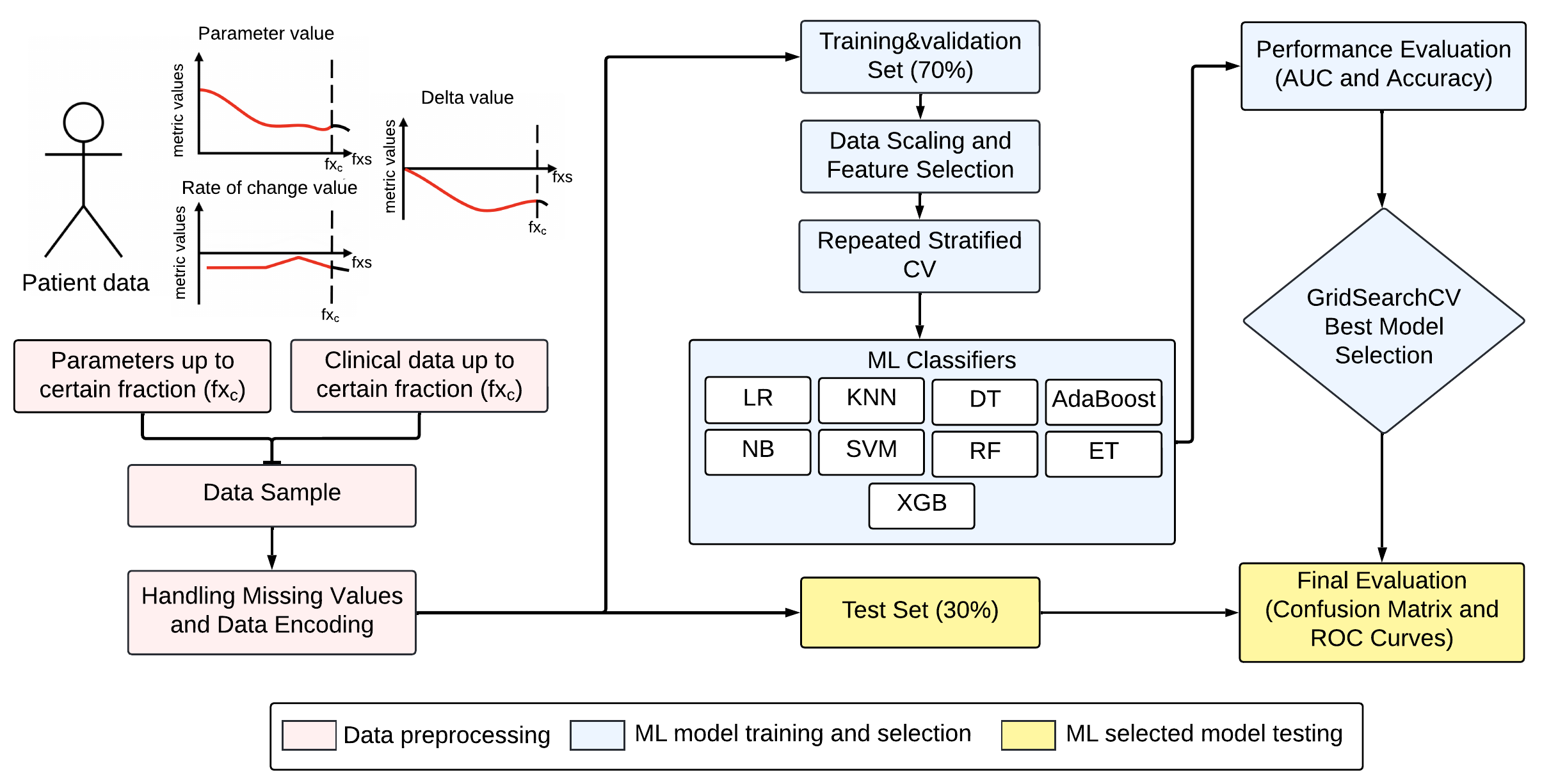}

\end{figure}
\subsubsection{Delta analysis}

Relative differences between the metric values at the time of the CT sim (fraction 0) and a fraction of interest were also considered in our analysis, since these values are frequently reported in literature \cite{El-Sayed2011}. These quantities were referred to as the Delta ($\Delta$) of each metric and were defined as $\Delta{p_{fx}}= \frac{p_{fx}-p_{0}}{p_{0}}$. Where $p_{fx}$ is the metric's value at the fraction of interest ($fx$) and $p_{0}$ is the metric's value at fraction 0 (CT sim). If the values of the metric at fraction 0 are zero, the $\Delta$ is instead calculated using the values of fraction 1. A similar approach has been employed in other studies, such as Delta radiomics for outcome prediction \cite{Xie2022-md, Cousin2023-of}. Histograms of the $\Delta$ values, representing the amount of anatomical change by the time of the replanning request, were generated to quantify the changes across the replanned patient sample.
\vspace{-3mm}

\subsection{ML models to predict if replanning will be needed}

After arranging and collecting the various metrics and clinical data, including the toxicity grades and patient weight at each fraction, missing values were identified and handled. Data were missing when the clinical questionnaires were not answered at every fraction or when not all patients had a CBCT taken at every fraction.

To handle missing values for individual patient paths (time series), three techniques were used: (1) filling the gap with the previous value in the series (in the case of toxicity grading), (2) filling the gap using linear interpolation (in the case of anatomical and geometrical metrics), and (3) filling the gap with the average of the data at a certain time point of the series. Lastly, categorical data, including TNM staging, sex, p16 status, smoking history, type of concomitant therapy, and HNC subsites, were handled using one-hot encoding. Ordinal encoding was used to preserve the hierarchy of the labels used for the overall cancer stage.  

Figure \ref{ML_flow} presents the pipeline used for the development and analysis of ML models. To build ML models as a prediction tool to identify which patients would likely need replanning, nine classifiers were evaluated: Logistic Regression (LR), KNearestNeighbors (KNN), Decision Tree (DT), AdaBoost, Naive Bayes (NB), Support Vector Machine (SVM), Random Forest (RF), ExtraTree (ET), and eXtreme Gradient Boost (XGB). Three specific fraction models were built using data up to and including fraction 5, fraction 10, and fraction 15, corresponding to the first, second, and third weeks of treatment. We selected these three timepoints since previous research found that patients experience anatomical variations from the first week that persist throughout treatment, with the most noticeable changes occurring mid-treatment \cite{Delaby2023-or, Avgousti2022-fq}. Additionally, patients may benefit from replanning before the third week of treatment \cite{Gan2024-xr}.

We split the data into 70\%-30\% training\&validation and test sets. Only the information available up to and including the fraction of interest was considered. To simulate real-world situations where the RT fraction at which the patient is being treated is the one we would like to evaluate, we filtered the training\&validation-testing split to include only the patients who had a request for replanning at or after the fraction for which the specific-fraction model is built.

To reduce the impact of data skewness or outliers \citep{Kuhn2019-yj}, the method StandardScaler from \textit{scikit-learn} was used for data scaling. To avoid overfitting, feature selection (FS) was performed using both filter and wrapper methods, specifically SelectKBest and RFE, respectively. Since the RFE technique uses a classifier algorithm, in our case, we used RandomForest as suggested and used by Chen et al. (2018)\cite{Chen2018-ie}, Darst et al. (2018)\cite{Darst2018-fp}, and Wang et al. (2022) \cite{Fan2022-dp}. All available data up to each fraction-specific time point were used as input for the FS techniques, allowing the algorithm to freely choose among the number of features $i$. The $i$ number was set to the number of metrics (43) plus the number of clinical characteristics and toxicities (21), for a total of $i = 64$ features. A subset comprising half of the total of inputs ($i =32$) was also tested to explore options with fewer degrees of freedom. FS techniques were only applied to the training set. 

A Repeated Stratified K-fold Cross Validation (CV) \cite{Hadjiiski2023-uq} method was used during the training and validation procedure to maintain the class proportion using $k = 5$ with ten repetitions. The performance of the ML models was evaluated using accuracy and AUC score. In each CV loop, performance metrics were obtained, along with their average, standard deviation and 95\% confidence interval. A GridSearchCV hypermetric tuning was done for the best-scoring models (based on the highest average AUC values) to improve performance. After model selection, the test set was used to estimate the final performance of the model, using the ROC curve, AUC, accuracy, and confusion matrix.
\vspace{-3mm}

\section{Results}
\vspace{-5mm}

\subsection{Study sample}
The study sample comprised various sites of diagnosis: oropharynx (44.7\%, N=67), oral cavity (52.0\%, N=39), larynx (28.0\%, N=21), nasopharynx (14.7\%, N=11), sinus (2.0\%, N=3), nasal cavity (1.3\%, N=2), hypopharynx (1.3 \%, N=2), and others (3.33\%, N=4), which included diagnoses related to the pharynx (N=1), cervical lymph nodes (N=1), and malignant neoplasm without specification of site in the head, face and neck region (N=2). This dataset consisted of male (72.0\%, N=108) and female (28.0\%, N=42) patients who were treated with total prescribed doses of 70 Gy (83.33\%, N=100) in 35 fractions, 60 Gy (10.83\%, N=13) in 30 fractions, and 66 Gy (5.83\%, N=7) in 33 fractions to the PTV, with curative intent. 

The treatment regimens included RT alone (34.0\%, N=51) as well as combined with systemic therapies, such as chemotherapy (62.0\%, N=93), targeted therapy (2.0\%, N=3), and immunotherapy (2.0\%, N=3). The T stage patient distribution corresponded to T4 (36.7\%, N=55), T3 (26.0\%, N=39), T2 (18.7\%, N=28), T1 (14.7\%, N=22), T0 (1.3\%, N=2), TX (2.0\%, N=3), and one unknown case (0.7\%, N=1). The N stage distribution was N3 (9.3\%, N=14), N2 (45.3\%, N=68), N1 (16\%, N=24), N0 (28.7\%, N=43), and unknown cases (0.7\%, N=1). The M stage distribution was M0 (92.0\%, N=138), M1 (4.7\%, N=7), MX (2.0\%, N=3), and unknown cases (1.3\%, N=2). The characteristics of the patient sample, separated into replanned and non-replanned patients, are presented in Table A in the Supplementary Material document.

\begin{table}[h!]
\captionv{12}{}{\justifying{Univariate analysis results for each created geometrical metric.
\label{TableParam2}}}
\vspace{-9mm}
\large
\begin{adjustbox}{width=1.01\textwidth}
\begin{tabular}{llll|llll}\\
\specialrule{2.pt}{0.5pt}{0.5pt}
\textbf{Metric} & \textbf{slopes fx} & \textbf{avg. AUC} & \textbf{$\Delta$}& \textbf{Metric} & \textbf{slopes fx} & \textbf{avg. AUC} & \textbf{$\Delta$}\\
\hline
\vspace{0.1mm}
\textit{(1) Body-related metrics}  &   &  & & \makecell*[l]
{$M_{avg}$} &   $7\rightarrow25$& 0.72 [0.63-0.79] &-4.7\%\\ 
\vspace{-1.5mm}
\makecell*[l]
{$V_{Body}$} &   $6\rightarrow25$& 0.75 [0.69-0.83] &-3.9\% &
\makecell*[l]{$\sigma_{M}$} &  ns & - &0.8\%\\
\vspace{-1.5mm}
\makecell*[l]{$CD_{Body}$} &3,$7\rightarrow25$ &0.77 [0.72-0.84] & 38.7\%& & & &\\
\cline{5-8}
\vspace{-1.5mm}
\makecell*[l]{$HD_{Body}$}& $12\rightarrow25$
&0.73 [0.64-0.80] &12.4\% & \multicolumn{4}{|l}{\textit{(5) Neck-related metrics}} \\
\makecell*[l]{$D_{Body}$} &9,$13\rightarrow25$ &0.75 [0.70-0.81] & 14.4\%&\makecell*[l]{$V_{Neck}$}& $6\rightarrow25$ &0.74 [0.66-0.80] & -6.1\%\\
\makecell*[l]{$\tilde{D}_{Body}$} & $9\rightarrow25$&0.80 [0.74-0.85] & 17.9\%& \makecell*[l]{$CD_{Neck}$} &$6\rightarrow25$ &0.78 [0.72-0.83] & 62.8\% \\
\makecell*[l]{$\bar{D}_{Body}$} & $7\rightarrow25$&0.80 [0.74-0.85] & 16.8\%&\makecell*[l]{
$HD_{Neck}$}& $21\rightarrow25$&0.71 [0.63-0.78] & 22.6\% \\
\cline{1-4}\multicolumn{4}{l|}{\textit{(2) Treatment mask-related metrics}} & \makecell*[l]{$D_{Neck}$}& $15,18\rightarrow25$&0.69 [0.61-0.77] &22.3\%\\
\makecell*[l]{$max\{B_{mask}\}$} &  ns & - &7.9\% & \makecell*[l]{$\tilde{D}_{Neck}$}& $14\rightarrow25$&0.76 [0.69-0.83] &34.4\%\\ 
\makecell*[l]{$\bar{B}_{mask}$} & $4\rightarrow25$&0.74 [0.66-0.81]& 30.2\% &\makecell*[l]{$\bar{D}_{Neck}$}& $14\rightarrow25$&0.77 [0.69-0.84]&30.9\%\\
\makecell*[l]{$\sigma_{{B}_{mask}}$} & ns & - & 5.2\% &\makecell*[l]{$R^{3D}_{min}$} & $8\rightarrow25$&0.69 [0.63-0.78]&-6.6\%\\
\makecell*[l]{$V^{air}_{Body-to-mask}$}& $6\rightarrow25$ &0.72 [0.64-0.79] & 26.5\% &\makecell*[l]{$R^{3D}_{max}$} & $19\rightarrow25$&0.68 [0.61-0.76]&-1.0\%\\
\cline{1-4}
\textit{(3) PTV-related metrics}     &   &  & &\makecell*[l]{
$R^{3D}_{avg}$}& 4,$6\rightarrow25$&0.78 [0.75-0.84]&-4.3\%\\
\makecell*[l]{$x^{PTV}_{min}$} & $6\rightarrow25$&0.77 [0.70-0.82] & -204.1\% &\makecell*[l]{
$\varphi^{3D}_{R}$}& ns&-&6.3\%\\
\makecell*[l]{$x^{PTV}_{max}$}  & ns &- &1.0\%&\makecell*[l]{$A^{2D}_{avg}$} & $6\rightarrow25$&0.75 [0.69-0.82]&-5.4\%\\
\makecell*[l]{$x^{PTV}_{avg}$} &4,$6\rightarrow25$ &0.76 [0.70-0.84] & -12.5\%& \makecell*[l]{$SA_{Neck}$}&$6\rightarrow25$&0.69 [0.64-0.75]&-0.7\%\\
\makecell*[l]{$x^{PTV}_{med}$} & $6\rightarrow25$&0.76 [0.68-0.83] & -15.5\%& \makecell*[l]{$C_{Neck}$}& ns &-&13.2\%\\
\cline{5-8}
\makecell*[l]{$x^{PTV}_{std}$} & ns&-&2.2\%& \multicolumn{4}{|l}{\textit{(6) Submandibular-related metrics}} \\
\makecell*[l]{$VI_{PTV}$} &ns &- & -8.0\%&
\makecell*[l]{$A_{sub}$} & $3\rightarrow25$&0.75 [0.70-0.81]&-5.2\%\\
\makecell*[l]{$VO_{PTV}$} & $4\rightarrow25$&0.70 [0.63-0.77]&-&\makecell*[l]{$R^{sub}_{min}$} & $6\rightarrow8,10\rightarrow25$&0.68 [0.61-0.77]&-6.0\%\\
$VI_{PTV}:LV_{Body}$ & 1,3,$8\rightarrow25$& 0.79 [0.73-0.85]&326.0\%&\makecell*[l]{$R^{sub}_{max}$} &  $16\rightarrow25$&0.67 [0.59-0.75]&-0.6\%\\
\makecell*[l]{$VO_{PTV}:LV_{Body}$} &$4\rightarrow25$ &0.69 [0.62-0.76] & -&\makecell*[l]{$R^{sub}_{avg}$} & $6\rightarrow25$ &0.75 [0.68-0.81]&-2.6\%\\
\cline{1-4}
\multicolumn{4}{l|}{\textit{(4) Mandible-related metrics}} &$\varphi^{2D}_{R^{sub}}$ & $18\rightarrow19$&0.65 [0.58-0.73]&6.4\%\\
\makecell*[l]{$M_{min}$} & $9\rightarrow25$&0.67 [0.60-0.75]&-24.8\%&
\makecell*[l]{$l^{sub}_{y}$}& 6,$8\rightarrow25$&0.70 [0.63-0.77]&-1.6\%\\
\makecell*[l]{$M_{med}$} & $10\rightarrow25$&0.71 [0.61-0.79]&-3.6\%&$l^{sub}_{x}$ & $6\rightarrow10$,$13\rightarrow25$&0.68 [0.61-0.75]&-2.8\%\\
\specialrule{2.pt}{0.5pt}{0.5pt}
\end{tabular}
\end{adjustbox}
\justify
\footnotesize{{\fontfamily{lmss}\selectfont{
Where the 'slopes fx' (with fx: fraction) columns indicate the fractions over which the results of the MWU tests applied to the slope calculations of the metrics were statistically significantly different (\textit{p-value}$<$0.05) for the replanned and non-replanned populations. ns: no significant results (\textit{p-value}$>$0.05). The $\rightarrow$ symbol indicates a range of fractions from a starting point to an ending point. The 'avg. AUC' is the average AUC for these selected fractions AUC, along with its 95$\%$ confidence interval.}}}
\bigskip
\end{table}

In this sample, the majority of the reported reasons for replanning were weight loss (70.7\%, N=53), followed by the PTV being in the air (10.7\%, N=8),  change in skin separation (13.3\%, N=10), skin reaction or swelling (10.7\%, N=8), mask looseness (10.7\%, N=8), and change in target volume (8.0\%, N=6). The most frequent fraction in which patients had a replanning request was 15, and the average fraction was 13. Note that several patients had more than one reason for replanning recorded in their charts.

\subsection{Univariate analysis}

Using our automatic extraction pipeline, all 43 metrics were calculated for the 150 patients in our study sample. This took an average of 25 minutes per patient using our computational setup. The results of the MWU test, as applied to the rate of change up to each fraction for each metric, are presented in Table \ref{TableParam2} (fractions at which \textit{p-value}$<$0.05). The average AUC values for the rate of change in the fractions that showed differences are also presented, along with the 95\% C.I. Additionally, the amount of change in each metric (using the delta analysis approach) that replanned patients experienced at the fraction at which the replanning was requested is presented with the symbol $\Delta$. The best univariate results for each category of metrics (based on the closest AUC values to 1) are graphically presented in Figure B of the Supplementary Material. In the case of the patient's rate of weight loss, the difference between the two groups was not notable until fraction 18, with an average AUC of 0.67 [0.58-0.75]. See Figure C in the Supplementary Material for more details.

\subsection{Machine learning analysis}

A total of 150 HNC patients were used for the ML analysis, corresponding to 75 replanned and 75 non-replanned patients. The split (70\%/30\%) of the data into training\&validation, and test sets corresponded to 105 patients and 45 patients, respectively. 105 patients (mean age of 62.9 years, 72 males and 33 females, 76 smokers, and 43 p16 positive) were included for training and CV, and 45 patients (mean age 63.6 years, 36 males and 9 females, 21 smokers, and 22 p16 positive) were included in the test set (see Tables B and C in the Supplementary Material). 

\begin{figure}[h!]
\includegraphics[width=0.325\linewidth]{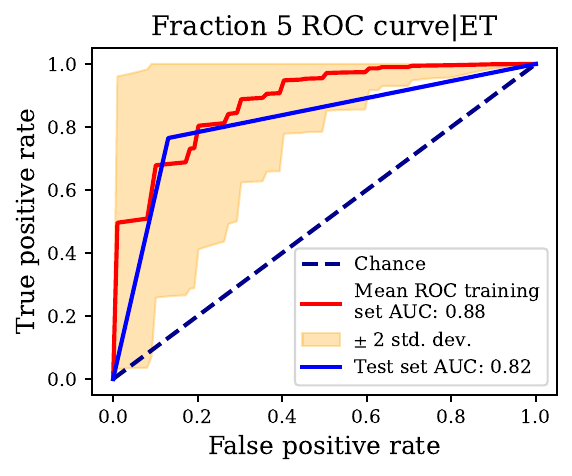}
\includegraphics[width=0.325\linewidth]{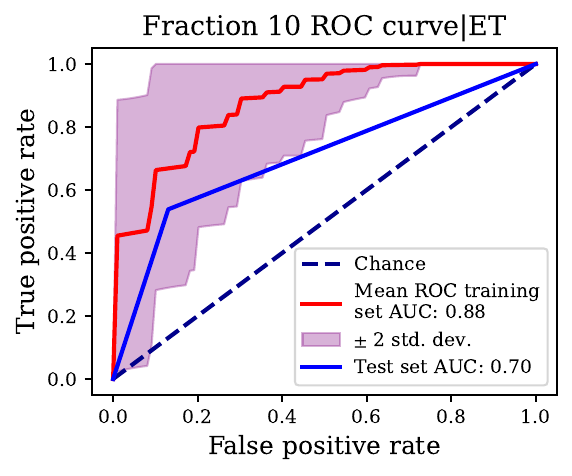}
\includegraphics[width=0.325\linewidth]{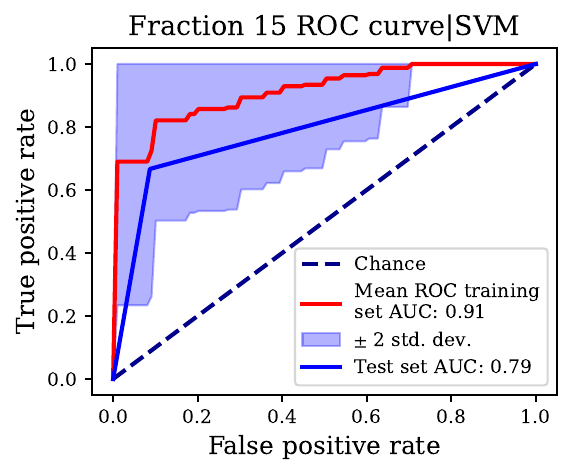}  
\centering
\includegraphics[width=0.28\linewidth]{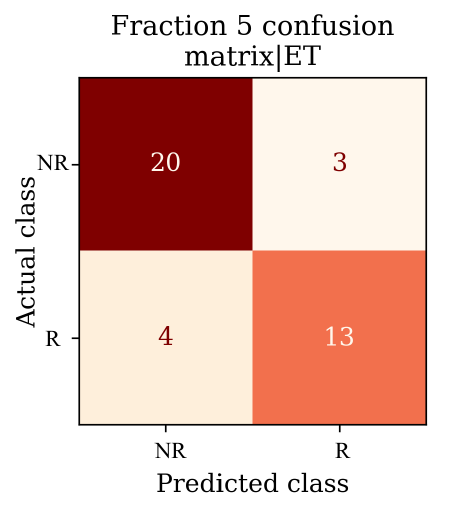}
\includegraphics[width=0.28\linewidth]{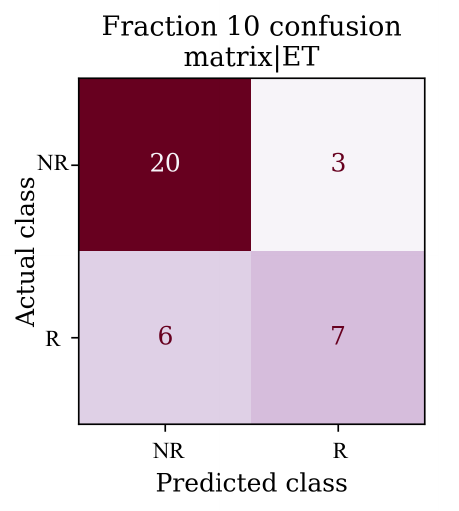}   \includegraphics[width=0.28\linewidth]{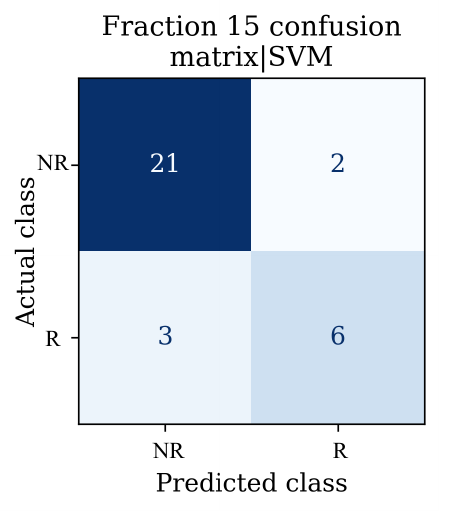}
\vspace{-13mm}
\captionv{12}{Short title - can be blank}{
\justifying{Fraction-specific model results. At the top are the ROC curves of the best model based on the training/validation CV results. At the bottom row, the confusion matrix results from the test set evaluation are shown. ET: ExtraTree Classifier and SVM: Support Vector Machine, with a sigmoid kernel. R stands for replanned and NR for non-replanned.}\label{ROCS}\vspace{3mm}}
\end{figure}

Models were built for three specific fractions (5, 10, and 15) to see if the metrics have the potential to predict the need for replanning using all the available data up to and including (but not after) the specified fraction. The models were tested using n=40, n=36, and n=32 patients, respectively. The results are presented in Figure \ref{ROCS}. Fraction 15-ML showed the lowest false positive and false negative cases. For more details (including selected geometrical metrics by the FS techniques and selected hyperparameters after the GridSearchCV tuning process) about each model, see Tables D, E, F, and G in the Supplementary Material. 
\vspace{-3mm}

\section{Discussion}
\vspace{-5mm}

Several authors have attempted to identify metrics that may influence RT replanning decisions in HNC patients. They focused on information available before treatment, such as TNM stage; overall cancer stage; and demographic information. According to Nuyts et al. (2024), Hu et al. (2018), and Chen et al. (2014) \cite{Nuyts2024-vy, Hu2018-md, Chen2014-fm}, replanned patients typically had, at the beginning of treatment, larger values in body weight, tumor volumes, and parotid gland volumes. However, no definitive method has been established to determine metrics related to the amount of weight loss or anatomical change in the HN region that requires replanning. Therefore, a tool that facilitates the identification of patients who would likely benefit from replanning would be very useful in the clinic.
\vspace{2mm}\\

In this research study, we developed a series of 43 metrics related to the geometrical shape of patients to describe the anatomical variation experienced by HNC patients during RT. Our work included the creation of an extraction tool called \textit{HNGeoNatomyX} that enables efficient assessment across fractions. The univariate and ML results presented demonstrate that these metrics can describe the anatomical evolution of replanned and non-replanned patients, showing their potential to differentiate between the two groups. Furthermore, the definition of the metrics, along with the \textit{HNGeoNatomyX} pipeline, will facilitate a potential analysis of $\Delta\%$ values that may help create a standard guideline or criteria for replanning evaluation, such as the ones proposed by Weppler et al. (2018)\cite{Weppler2018-nv} and Bak et al. (2022)\cite{Bak2022-bv}. Thus, statistical approaches based on ROC curves or histograms of the metric variation may be a good approach for establishing clinical threshold values \cite{Hassanzad2024-qi}.  

Additionally, the incorporation of the metrics in an ML model analysis showed promising results in predicting whether replanning is needed, which could facilitate the RT treatment workflow for HNC patients. These models offer the potential to help clinical personnel determine as early as possible which patients will likely need treatment replanning, allowing timely intervention and improved workflow efficiency. 

To the best of our knowledge, no other studies have done an ML analysis to predict RT replanning based on clinical data and geometrical metrics of the patient's shape over time. Given that our methodology includes information that radiation oncologists are familiar with in the clinic, the replanning prediction can be used for double verification. Regarding clinical implementation, once the models can be trained on a larger dataset and tested in an external cohort, this tool may potentially be integrated into treatment planning software to support decisions for RT treatment replanning. Including the automatic extraction pipeline in such software may help track important features and statistics over time, contributing to a standardized replan decision. Currently, this decision depends on individual radiation oncologists' and medical physicists' judgments, leading to larger variations in the decision.

We acknowledge some limitations in our study. One of the main limitations is that some non-replanned patients considered in our study may have actually needed to be replanned, but they were not due to undocumented logistical reasons, such as consideration of the number of remaining fractions. Some of these cases might be reflected in the false positive cases shown in the confusion matrix. To overcome this type of limitation, a comprehensive analysis of the patients initially labeled as non-replanned needs to be reassessed. This could be done on a case-by-case basis and/or using clustering techniques. Another limitation worth mentioning is that the metrics were extracted for all the fractions available simultaneously, which affects the body contour-cutting process to match the height of the bodies. Although this is correct for comparing the same region, to improve the input for ML models in a real-life scenario, this step should consider the contours up to the fraction to which the specific model corresponds. Finally, it is important to mention that currently our models might not be generalizable since, even though our study sample covered a wide range of HNC subsites, it did not have significant samples for certain cases (salivary glands, nasal cavity, and hypopharynx), which might not be enough to provide confident results. However, with our developed and tested workflow and our code available under an open-source license, an analysis with a larger, ideally multicentre dataset, should be possible.
\vspace{-3mm}

\section{Conclusions}
\vspace{-5mm}

We defined a number of geometrical metrics that describe the anatomical changes experienced by HNC patients during RT and demonstrated that they can be used to characterize and distinguish patients who will and will not require replanning. Our automatic metric calculation pipeline (\textit{HNGeoNatomyX}) and associated ML models offer the potential to help streamline HNC patient resource management and clinical workflow in RT.

\vspace{-5mm}
\section*{Code Availability}
\vspace{-5mm}

The automatic extraction pipelines for calculating the metrics and the semi-automatic treatment mask contouring algorithm that support the findings of this study are available under an open source license on kildealab GitHub in the HNGeoNatomyX section: \url{github.com/kildealab/HNGeoNatomyX}.

\vspace{-5mm}
\section*{Acknowledgements}
\vspace{-5mm}

This research was supported by the Rossy Cancer Network, the CREATE Responsible Health and Healthcare Data Science (SDRDS) grant from the Natural Sciences and Engineering Research Council, and the Research Institute of the MUHC (RI-MUHC). The authors acknowledge Ackeem Joseph, MSc; Aixa X. Andrade Hernandez, MSc; and Victor Matassa, BSc, for their valuable contributions in sourcing the data used in this study.

\vspace{-5mm}

\section*{Author Contribution Statement}
\vspace{-5mm}

\textbf{Conceptualization:} Odette Rios-Ibacache, James Manalad, Julia Khriguian, George Shenouda, and John Kildea; \textbf{Methodology and software:} Odette Rios-Ibacache and James Manalad; \textbf{Formal Analysis:} Odette Rios-Ibacache, Kayla O'Sullivan-Steven, Luc Galarneau, and John Kildea; \textbf{Data curation:} Odette Rios-Ibacache, Kayla O'Sullivan-Steben, and Emily Poon; \textbf{Manuscript writing:} Odette Rios-Ibacache and John Kildea; \textbf{Manuscript editing:} Odette Rios-Ibacache, John Kildea, James Manalad, Kayla O'Sullivan-Steben, Emily Poon, Luc Galarneau, Julia Khriguian, and George Shenouda; \textbf{Supervision:} John Kildea; \textbf{Funding acquisition:}  John Kildea. All authors have read and agreed to the published version of the manuscript.
\vspace{-5mm}

\section*{Conflict of Interest Statement}
\vspace{-5mm}

The authors have no conflicts of interest to declare.
\vspace{-5mm}
\section*{References}
\vspace{-5mm}

\footnotesize
\addcontentsline{toc}{section}{\numberline{}References}
\setstretch{0.9}

\bibliographystyle{./medphy.bst} 








\pagebreak

\normalsize
\cen{\sf {\Large {\bfseries Supplementary Material} 

Odette Rios-Ibacache$^{1}$,
James Manalad$^{1}$,
Kayla O'Sullivan-Steben$^{1}$,
Emily Poon$^{2}$,
Luc Galarneau$^{1,2}$,
Julia Khriguian$^{3}$,
George Shenouda$^{4}$ \&
John Kildea$^{1,2}$}

$^{1}$Medical Physics Unit, McGill University, Montreal, QC, Canada\\
$^{2}$Research Institute of the McGill University Health Centre, McGill University, Montreal, QC, Canada\\
$^{3}$Department of Radiation Oncology, Hôpital Maisonneuve-Rosemont, Montreal, QC, Canada. \\
$^{4}$Department of Radiation Oncology, McGill University, Montreal, QC, Canada.
\vspace{5mm}\\
Version typeset \today\\}
\renewcommand{\thetable}{\Alph{table}}
\renewcommand{\thefigure}{\Alph{figure}}
\setcounter{figure}{0}
\setcounter{table}{0}
\setcounter{section}{0}
\pagenumbering{roman}
\setcounter{page}{1}
\pagestyle{plain}

\vspace{-13mm}
\section{Introduction}

\begin{figure}[h]
\captionv{12}{}
{\justifying{Decision tree outlining the inclusion and exclusion criteria for the patient sample.}\label{Inclusion}}
\begin{center}
\includegraphics[width=1\linewidth]{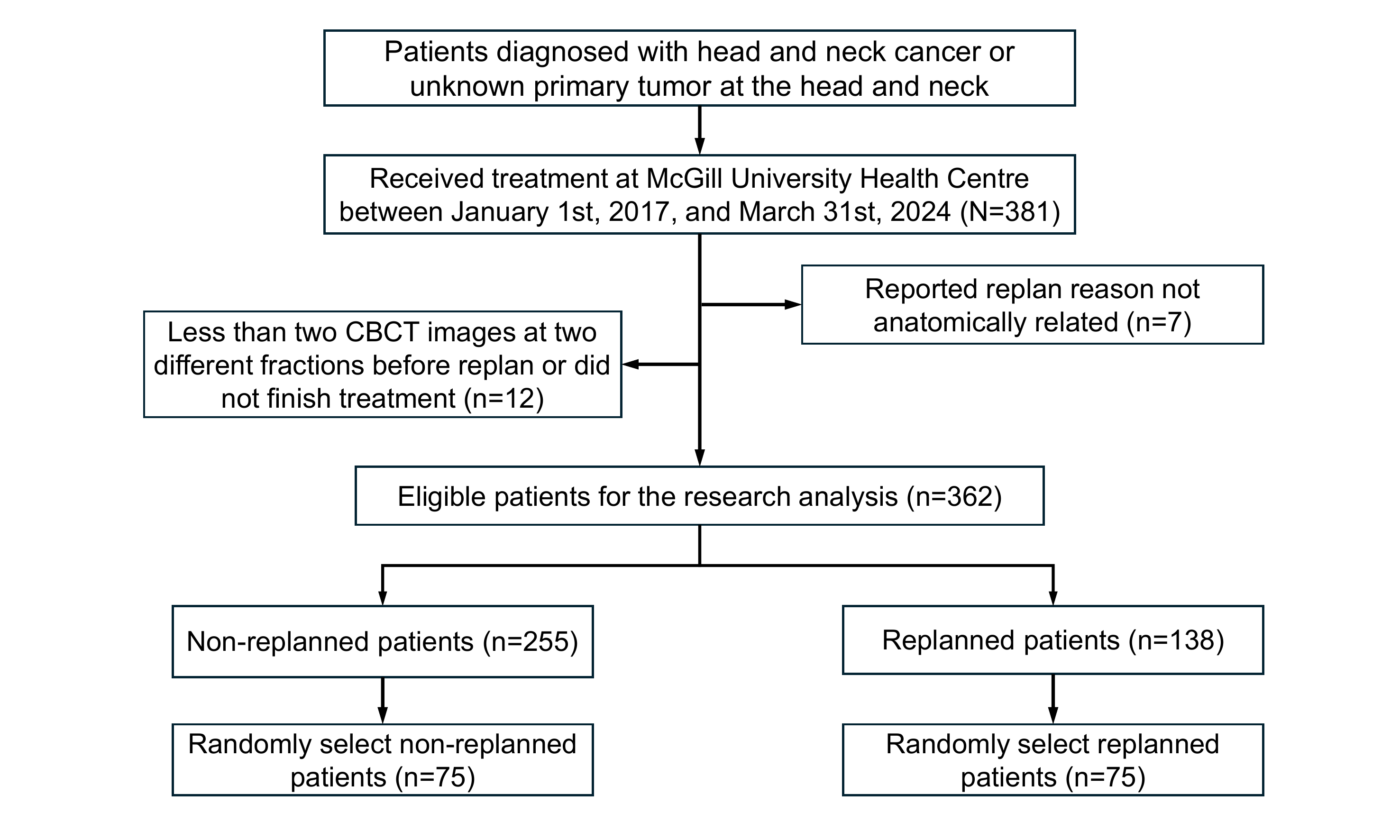} 
\end{center}
\end{figure}
\newpage

\section{Results}

Table A presents the characteristics of the patient sample used in this research study, categorized into replanned and non-replanned patients.

\vspace{5mm}
\begin{table*}[h!]
\captionv{12}{Short title - can be blank}{
\justifying{Characteristics of the patient sample used in this research project. R indicates the replanned patient category and NR the non-replanned patient category. *Other cancer sites correspond to malignant neoplasm \textbf{without specification of site}, of lymph nodes of the head, face, and neck or tonsils. $p$ is the p-value from a: Mann–Whitney U or b: Fisher's test. Statistically significant values are in bold.
\label{sample}}}
\vspace{-5mm}
\centering
\footnotesize
\begin{tabular}{>{\footnotesize}r>{\footnotesize}c>{\footnotesize}c>{\footnotesize}c|>{\footnotesize}r>{\footnotesize}c>{\footnotesize}c>{\footnotesize}c}
\hline
\hline
\multicolumn{1}{>{\footnotesize}l}{\textbf{Characteristic}} & \textbf{R}& \textbf{NR} & \textbf{\textit{p}}&\multicolumn{1}{>{\footnotesize}l}{\textbf{Characteristic}} & \textbf{R} & \textbf{NR} & \textbf{\textit{p}}\\ \hline
\multicolumn{1}{>{\footnotesize}l}{\textbf{Age}} &  &  & \multicolumn{1}{>{\footnotesize}l|}{ns$^{a}$}&
\multicolumn{1}{>{\footnotesize}l}{\textbf{N Stage}} &  &  &\multicolumn{1}{>{\footnotesize}l}{\textit{}} \\
\multicolumn{1}{>{\footnotesize}l}{} average (years) & 61.7 & 64.6& &\textbf{ N0} & \textbf{13 (17.3\%)} & \textbf{30 (40.0\%)} & \textbf{$<$0.05} \\
\multicolumn{1}{>{\footnotesize}l}{} range (years) & (38,84) & (41,85)& \multicolumn{1}{>{\footnotesize}l|}{} & N1 & 12 (16.0\%) & 12 (16.0\%)& ns$^{b}$ \\
\multicolumn{1}{>{\footnotesize}l}{\textbf{Sex}} & & & \multicolumn{1}{>{\footnotesize}l|}{} & N2 & 41 (54.7\%) & 27 (36.0\%) & ns$^{b}$\\
male & 56 (74.7\%) & 52 (69.3\%)&ns$^{b}$ & N3 & 8 (10.7\%) &  6 (8.0\%) & ns$^{b}$ \\
female & 19 (25.3\%) & 23 (30.7\%) & ns$^{b}$ & unknown & 1 (1.3\%) & 0 (0.0\%) & ns$^{b}$ \\
\multicolumn{1}{>{\footnotesize}l}{\textbf{Site}} & & & \multicolumn{1}{>{\footnotesize}l|}{} & \multicolumn{1}{>{\footnotesize}l}{\textbf{M Stage}} &  & & \textit{} \\
oropharynx & 36 (48.0\%) & 31 (41.3\%) & ns$^{b}$  & M0 &  66 (88.0\%) & 72 (96.0\%) & ns$^{b}$\\
oral cavity & 18 (24.0\%) &21 (28.0\%)& ns$^{b}$  & M1 &  5 (6.7\%)&2 (2.7\%)&  ns$^{b}$\\
larynx & 8 (10.7\%) & 13 (17.3\%) & ns$^{b}$ & MX & 3 (1.3\%) & 0 (0.0\%) & ns$^{b}$ \\
nasopharynx & 5 (6.7\%) & 6 (8.0\%)& ns$^{b}$  & unknown & 1 (1.3\%) & 1 (1.3\%) & ns$^{b}$\\
nasal cavity & 1 (1.3\%) & 1 (1.3\%) & ns$^{b}$ &  \multicolumn{1}{>{\footnotesize}l}{\textbf{Prescription dose}} &  &  & \textit{}  \\
hypopharynx & 2 (2.7\%) & 0 (0.0\%) & ns$^{b}$ &70 Gy/35 fxs & 55 (73.3\%) & 61 (81.3\%) & ns$^{b}$\\
salivary glands & 1 (1.3\%) & 0 (0.0\%) & ns$^{b}$& 66 Gy/33 fxs & 10 (13.3\%) & 7 (9.3\%)& ns$^{b}$\\
sinus & 0 (0.0\%) & 3 (4.0\%) & ns$^{b}$  & 60 Gy/30 fxs & 10 (13.3\%) & 7 (9.3\%) & ns$^{b}$\\
others* & 4 (5.3\%) & 0 (0.0\%) & ns$^{b}$ & \multicolumn{1}{>{\footnotesize}l}{\textbf{Systematic therapy}} &  &  & \\
\multicolumn{1}{>{\footnotesize}l}{\textbf{Cancer Stage}} &  &  &\multicolumn{1}{>{\footnotesize}l|}{} & RT alone & 25 (33.3\%) & 27 (36.0\%)& ns$^{b}$\\
I & 2 (2.7\%) & 3 (4.0\%) & ns$^{b}$ &  chemotherapy & 49 (65.3\%) & 44 (58.7\%)&  ns$^{b}$\\
II & 1 (1.3\%) & 1 (1.3\%) & ns$^{b}$& targeted therapy & 1 (1.3\%) & 3 (4.0\%) & ns$^{b}$\\
III & 9 (12.0\%) & 14 (18.7\%) & ns$^{b}$ & immunotherapy & 0 (0.0\%) & 1 (1.3\%) & ns$^{b}$\\
IV & 33 (44.0\%) &  26 (34.7\%) & ns$^{b}$ &  \multicolumn{1}{>{\footnotesize}l}{\textbf{p16 status}}&  & & \multicolumn{1}{>{\footnotesize}l}{}\\
 X & 29 (38.7\%) & 31 (41.3\%) &\multicolumn{1}{>{\footnotesize}l|}{} & positive & 36 (48.0\%) &  29 (38.7\%) \\
unknown & 1 (1.3\%) & 0 (0.0\%) & ns$^{b}$ & negative & 20 (26.7\%) & 17 (22.7\%) &ns$^{b}$ \\
\multicolumn{1}{>{\footnotesize}l}{\textbf{T stage}} & & &\multicolumn{1}{>{\footnotesize}l|}{}  & unknown &  19 (25.3\%) & 29 (38.7\%) & ns$^{b}$\\
T0 & 2 (2.7\%) & 0 (0.0\%) & ns$^{b}$  &  \multicolumn{1}{>{\footnotesize}l}{\textbf{Smoking history}} & & & \multicolumn{1}{>{\footnotesize}l}{}  \\
T1 & 12 (16.0\%) & 10 (13.3\%) & ns$^{b}$ &  smoker/smoked  & 54 (72.0\%) & 53 (70.7\%) & ns$^{b}$\\
T2 & 12 (16.0\%) & 16 (21.3\%) & ns$^{b}$ & never-smoker& 20 (26.7\%) & 20 (26.7\%) & ns$^{b}$\\
T3 & 16 (21.3\%) &  23 (30.7\%) & ns$^{b}$ & unknown & 1 (1.3\%) & 2 (2.7\%) & ns$^{b}$ \\
T4 & 29 (38.7\%) & 26 (34.7\%)  & ns$^{b}$ &  &  & & \\
TX & 3 (4.0\%) &  0 (0.0\%)  & ns$^{b}$ & &  & & \\
unknown & 1 (1.3\%) &  0 (0.0\%)  & ns$^{b}$ & &  & & \\
\hline 
\hline
\end{tabular}
\end{table*}

\vspace{5mm}

In Figure B are graphically presented, based on AUC values closest to 1, the best univariate results for each metric category. Generally, the median values of the longitudinal data distributions for the two classes (replanned and non-replanned) differ in the case of the body-related metrics. Replanned versus non-replanned presented significant differences.

\begin{figure*}[h!]
\captionv{12}{Short title - can be blank}{
\justifying{{Progression over time (over fractions) of the best results metrics based on AUC values closest to 1 for each metric category. Next to each metric’s longitudinal graph, are the results from the Mann-Whitney U test for their rate of change, for six different fractions. In red is the first fraction at which the p-value is constantly below 0.05. On the right is the amount of change of the metrics ($\Delta$\%), concerning fraction 0 (for $A_{sub}$, $\bar{B}_{mask}$, $x^{PTV}_{avg}$ and $M_{min}$) or fraction 1 (for $CD_{Neck}$ and $\bar{D}_{Body}$), distribution values by
the time of the request replan fraction.}}
\label{body_params}}

\includegraphics[width=0.32\linewidth]{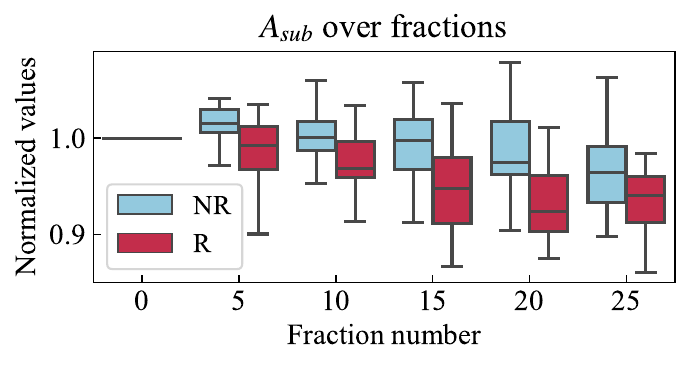}
\includegraphics[width=0.32\linewidth]{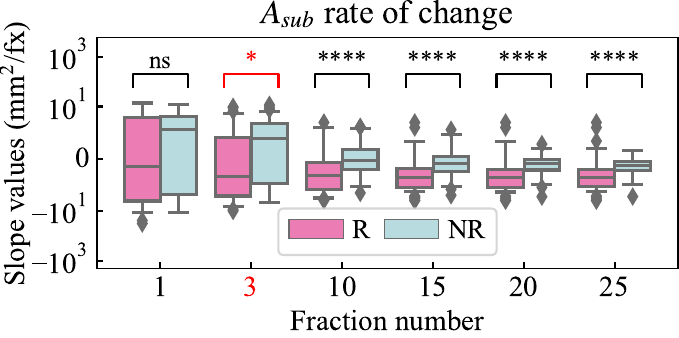}
\includegraphics[width=0.32\linewidth]{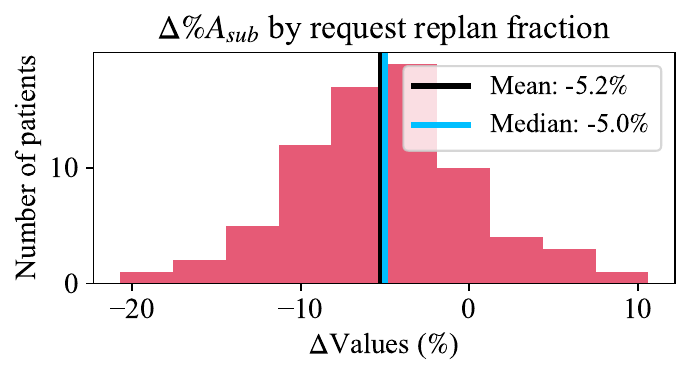}
\includegraphics[width=0.32\linewidth]{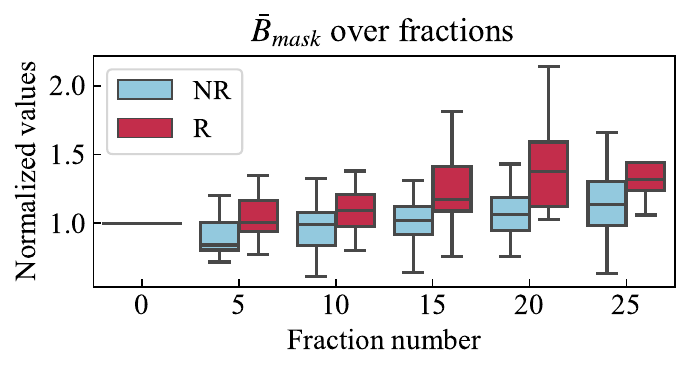}
\includegraphics[width=0.32\linewidth]{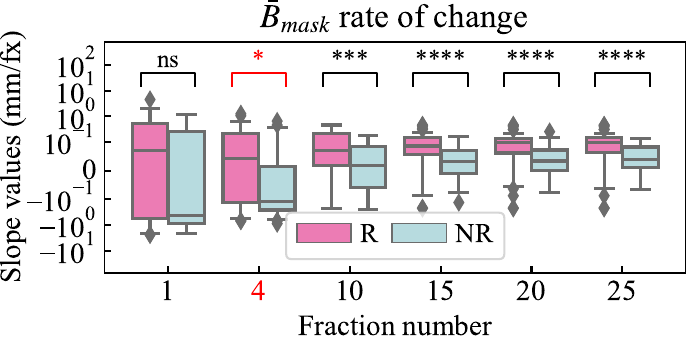}
\includegraphics[width=0.32\linewidth]{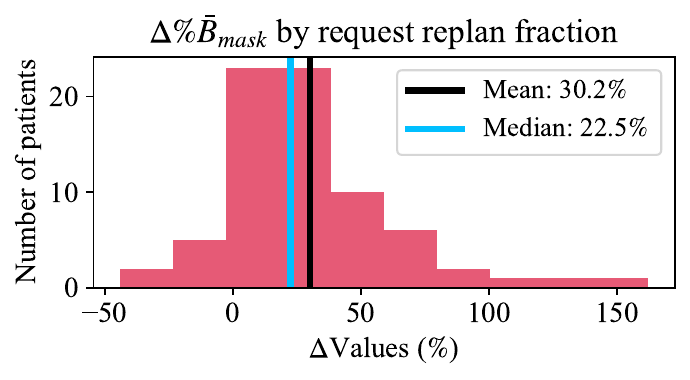}
\includegraphics[width=0.32\linewidth]{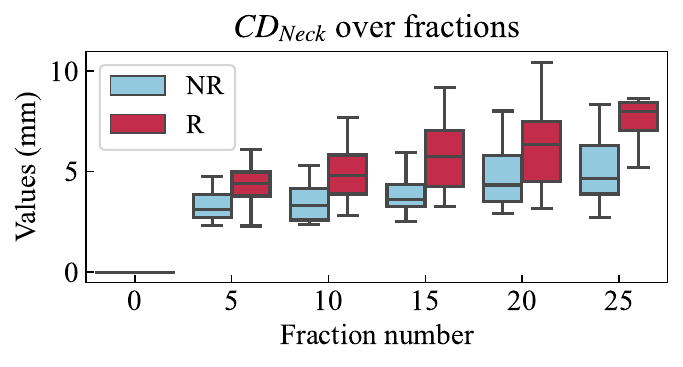}
\includegraphics[width=0.32\linewidth]{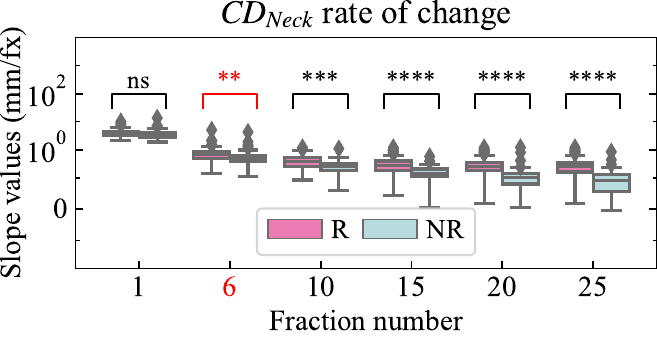}
\includegraphics[width=0.32\linewidth]{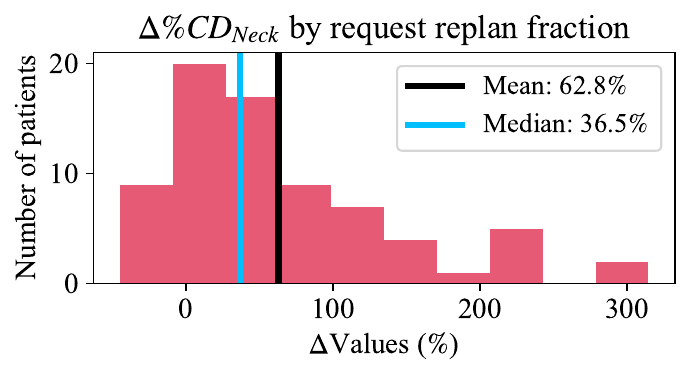}
\includegraphics[width=0.32\linewidth]{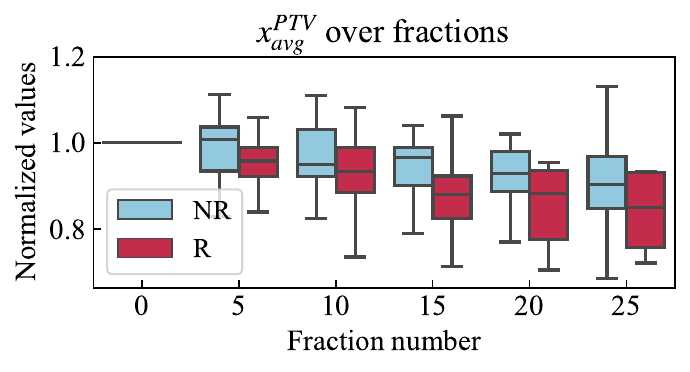}
\includegraphics[width=0.32\linewidth]{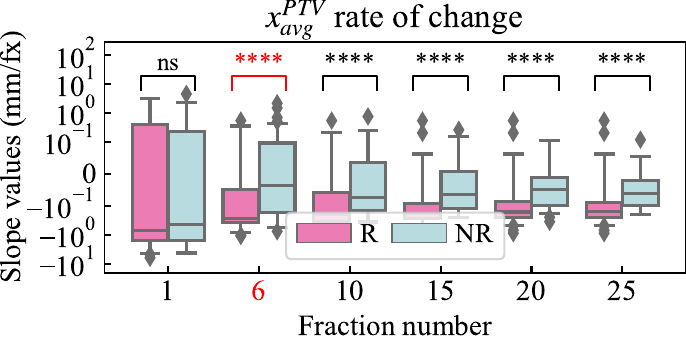}
\includegraphics[width=0.32\linewidth]{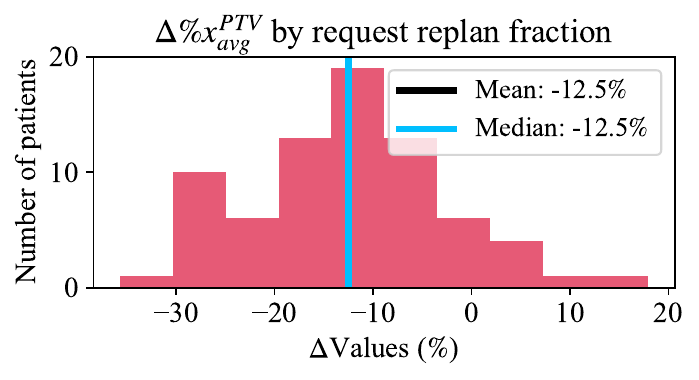}
\includegraphics[width=0.32\linewidth]{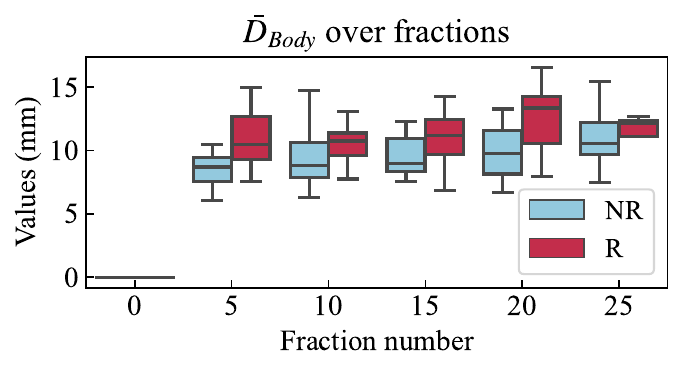}
\includegraphics[width=0.32\linewidth]{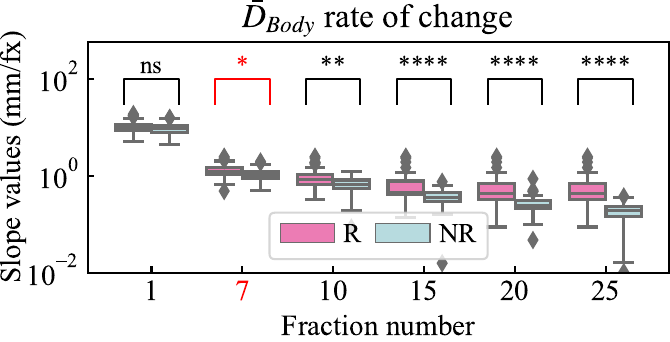}
\includegraphics[width=0.32\linewidth]{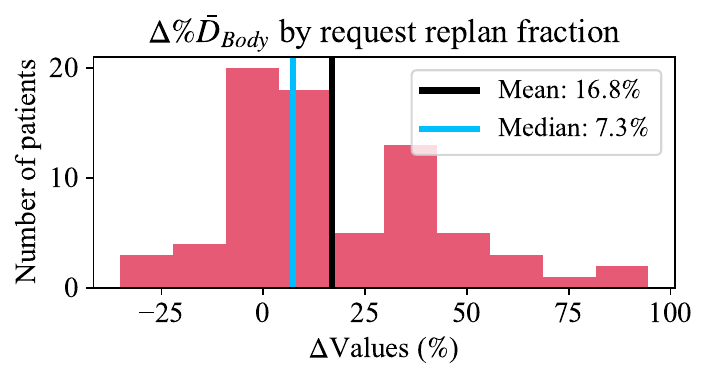}
\includegraphics[width=0.32\linewidth]{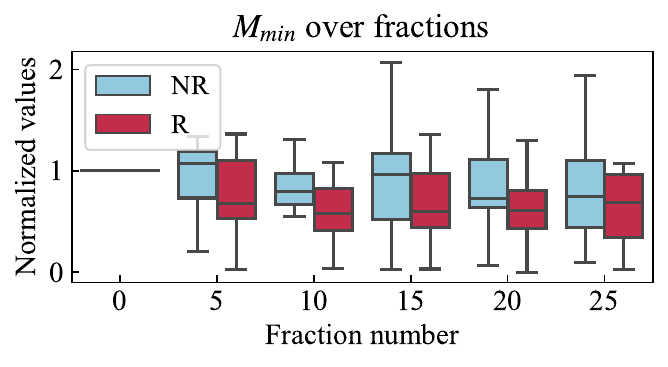}
\includegraphics[width=0.32\linewidth]{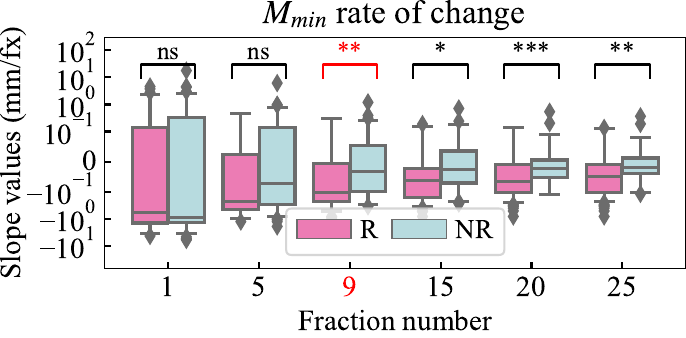}
\includegraphics[width=0.32\linewidth]{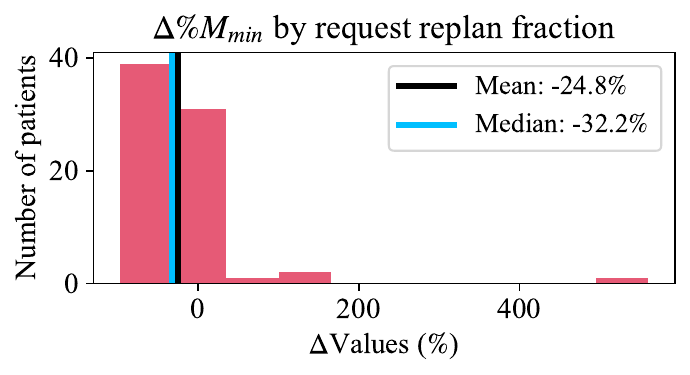}
\includegraphics[width=1\linewidth]{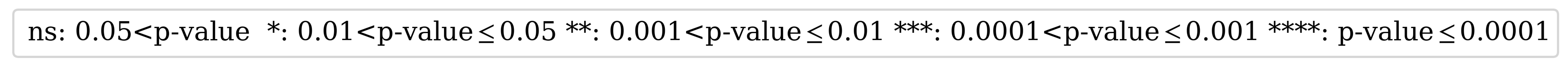}
\end{figure*}

\begin{figure*}[h]
\captionv{12}{Short title - can be blank}{
\justifying{{On the left is the graph showing normalized weight values over time for both replanned and non-replanned patients. In the middle panel are the results from the Mann-Whitney U test for the rate of change in weight across six different fractions. In red is the first fraction at which the p-value is constantly below 0.05. On the right is the amount of change of the weight ($\Delta$\%) concerning the fraction 0 distribution values, by the time of the request replan fraction.\vspace{5mm}}}
\label{weightss}}
   \centering 
\includegraphics[width=0.32\linewidth]{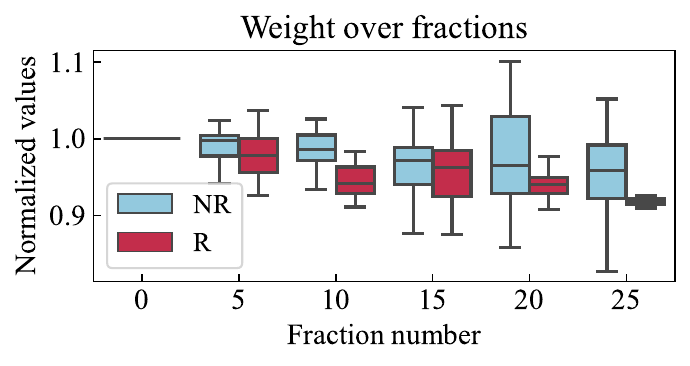}    \includegraphics[width=0.32\linewidth]{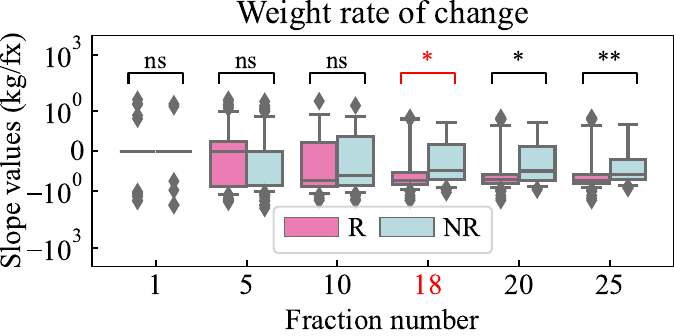}
\includegraphics[width=0.32\linewidth]{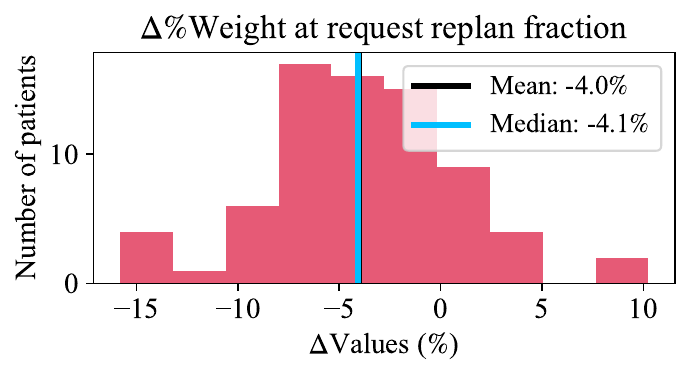}
\includegraphics[width=1\linewidth]{supplementary/legend.pdf} 
\end{figure*}

\section{Machine Learning}

\begin{table*}[h!]
\captionv{12}{Short title - can be blank}{
\justifying{Patient characteristics in the training \& validation, and test sets. The training\&validation set corresponds to 105 patients, while the testing set corresponds to 45 patients. fx is an abbreviation for fraction, R for replanned, and NR for non-replanned.  $p$ is the p-value from a: Mann–Whitney U or b: Fisher's test.
\label{split1}}}
\vspace{-5mm}
\centering
\begin{tabular}{>{\footnotesize}r>{\footnotesize}c>{\footnotesize}c>{\footnotesize}c>{\footnotesize}c>{\footnotesize}c}
\hline
\hline
\multicolumn{1}{>{\footnotesize}l}{\textbf{Characteristics}} & \multicolumn{2}{>{\footnotesize}l}{\textbf{Training\&validation set (n = 105)}} & \multicolumn{2}{>{\footnotesize}l}{\textbf{Testing set (n = 45)}} & \textit{\textbf{p}} \\ \hline
\multicolumn{1}{>{\footnotesize}l}{} & \textbf{R} & \textbf{NR} & \textbf{R} & \textbf{NR} \\
{\textbf{N}} & 53 & 52 & 22 & 23\\
\multicolumn{1}{>{\footnotesize}l}{\textbf{Age}} &  &  &  &  & ns$^{a}$\\
average (years) & 61.4 & 64.6 & 62.5 & 64.6 &\\
range (years) & (41, 79) & (41,85) & (38, 84) & (41,84) \\
\multicolumn{1}{>{\footnotesize}l}{\textbf{Sex}} & \multicolumn{1}{>{\footnotesize}l}{} & \multicolumn{1}{>{\footnotesize}l}{} & \multicolumn{1}{>{\footnotesize}l}{} & \multicolumn{1}{>{\footnotesize}l}{} &  \\
male & 39 (36.2\%) & 33 (34.3\%) & 17 (40.0\%) & 19 (35.6\%) & ns$^{b}$\\
female & 14 (14.3\%) & 19 (15.2\%) & 5 (8.9\%) & 4 (15.6\%) & ns$^{b}$\\
\multicolumn{1}{>{\footnotesize}l}{\textbf{Request replan fx}} & \multicolumn{1}{>{\footnotesize}l}{} & \multicolumn{1}{>{\footnotesize}l}{} & \multicolumn{1}{>{\footnotesize}l}{} & \multicolumn{1}{>{\footnotesize}l}{} & ns$^{a}$\\
average & 14 & - & 12 & - \\
range & (2, 26) & - & (1, 23) & - \\
\multicolumn{1}{>{\footnotesize}l}{\textbf{Site}} & \multicolumn{1}{>{\footnotesize}l}{} & \multicolumn{1}{>{\footnotesize}l}{} & \multicolumn{1}{>{\footnotesize}l}{} & \multicolumn{1}{>{\footnotesize}l}{} \\
oropharynx & 26 (49.1\%) & 20 (38.5\%) & 10 (45.5\%) & 11 (47.8\%) & ns$^{b}$\\
oral cavity & 13 (24.5\%) & 13 (25.0\%) & 5 (22.7\%) & 8 (34.8\%) & ns$^{b}$\\
larynx & 6 (11.3\%) & 10 (19.2\%) & 2 (9.1\%) & 3 (13.0\%) &  ns$^{b}$\\
nasopharynx & 4 (7.5\%) & 6 (11.5\%) & 1 (4.5\%) & 0 (0.0\%) & ns$^{b}$\\
nasal cavity & 1 (1.9\%) & 1 (1.9\%) & 0 (0.0\%) & 0 (0.0\%) & ns$^{b}$\\
hypopharynx & 0 (0.0\%) & 0 (0.0\%) & 2 (9.1\%) & 0 (0.0\%) & ns$^{b}$ \\
sinus &  0 (0.0\%)& 2 (3.8\%) & 0 (0.0\%) & 1 (4.3\%) & ns$^{b}$\\
salivary glands & 1 (1.9\%) & 0 (0.0\%)&  0 (0.0\%)& 0 (0.0\%) & ns$^{b}$\\
others & 2 (3.8\%) & 0 (0.0\%) & 2 (9.1\%) & 0 (0.0\%) & ns$^{b}$\\
\multicolumn{1}{>{\footnotesize}l}{\textbf{Prescription dose}} & \multicolumn{1}{>{\footnotesize}l}{} & \multicolumn{1}{>{\footnotesize}l}{} & \multicolumn{1}{>{\footnotesize}l}{} & \multicolumn{1}{>{\footnotesize}l}{} \\
70 Gy/35 fxs & 41 (77.4\%) & 44 (84.6\%) & 14 (63.6\%) & 17 (73.9\%) & ns$^{b}$\\
66 Gy/33 fxs & 7 (13.2\%) & 4 (7.7\%) & 3 (13.6\%) & 3 (13.0\%) & ns$^{b}$\\
60 Gy/30 fxs & 5 (9.4\%) & 4 (7.7\%) & 5 (22.7\%) & 3 (13.0\%) & ns$^{b}$\\
\multicolumn{1}{>{\footnotesize}l}{\textbf{Smoking history}} & \multicolumn{1}{>{\footnotesize}l}{} &  & \multicolumn{1}{>{\footnotesize}l}{} &  \\
smoker & 38 (71.7\%) & 38 (73.1\%) & 16 (72.7\%) & 15 (65.2\%) & ns$^{b}$\\
non-smoker & 14 (26.4\%) & 13 (25.0\%) & 6 (27.3\%) & 7 (30.4\%) & ns$^{b}$\\
unknown &   1 (1.9\%) & 1 (1.9\%)    &   0 (0.0\%) & 1 (4.3\%) & ns$^{b}$\\
\multicolumn{1}{>{\footnotesize}l}{\textbf{Systemic therapy}} &  & \multicolumn{1}{>{\footnotesize}l}{} &  & \multicolumn{1}{>{\footnotesize}l}{} \\
RT alone &  17 (32.1\%)& 16 (30.8\%)& 8 (35.4\%)& 11 (47.8\%)& ns$^{b}$\\
chemotherapy & 35 (66.0\%)& 34 (65.4\%)& 14 (63.6\%)& 10 (43.5\%)& ns$^{b}$ \\ 
targeted therapy & 1 (1.9\%)& 1 (1.9\%)&  0 (0.0\%)& 2 (8.7\%) & ns$^{b}$\\
immunotherapy & 0 (0.0\%)& 1 (1.9\%)& 0 (0.0\%) & 0 (0.0\%) & ns$^{b}$\\
\multicolumn{1}{>{\footnotesize}l}{\textbf{p16 status}} & \multicolumn{1}{>{\footnotesize}l}{} & \multicolumn{1}{>{\footnotesize}l}{} & \multicolumn{1}{>{\footnotesize}l}{} &  \\
positive & 26 (49.1\%) & 17 (32.7\%) & 10 (45.5\%) & 12 (52.2\%) & ns$^{b}$\\
negative & 15 (28.3\%) & 14 (26.9\%) & 5 (22.7\%) & 3 (13.0\%) & ns$^{b}$\\
unknown & 12 (22.6\%) & 21 (40.4\%) & 7 (31.8\%) & 8 (34.8\%) & ns$^{b}$\\ \hline
\hline
\end{tabular}
\captionsetup{justification=justified,font=footnotesize}
\vspace{5mm}
\end{table*}

\begin{table*}[h!]
\captionv{12}{Short title - can be blank}{
\justifying{Patient's cancer overall stage and TNM staging in the training\&validation, and test sets. R stands for replanned, and NR stands for non-replanned.  $p$ is the p-value from a: Mann–Whitney U or b: Fisher's test.
\label{demo2}}}
\vspace{-5mm}
\centering
\begin{tabular}{>{\footnotesize}r>{\footnotesize}c>{\footnotesize}c>{\footnotesize}c>{\footnotesize}c>{\footnotesize}c}
\hline
\hline
\multicolumn{1}{>{\footnotesize}l}{\textbf{Characteristics}} & \multicolumn{2}{>{\footnotesize}l}{\textbf{Training and validation set (n = 105)}} & \multicolumn{2}{>{\footnotesize}l}{\textbf{Testing set (n = 45)}} \\ \hline
\multicolumn{1}{>{\footnotesize}l}{} & \textbf{R} & \textbf{NR} & \textbf{R} & \textbf{NR} \\
\multicolumn{1}{>{\footnotesize}l}{\textbf{Cancer stage}} &  & \multicolumn{1}{>{\footnotesize}l}{} &  & \multicolumn{1}{>{\footnotesize}l}{} \\
I & 2 (3.8\%) & 3 (5.8\%) & 0 (0.0\%) & 0 (0.0\%) &  ns$^{b}$\\
II & 1 (1.9\%) & 1 (1.9\%) & 0 (0.0\%) & 0 (0.0\%) & ns$^{b}$ \\
III & 6 (11.3\%) & 10 (19.2\%) & 3 (13.6\%) & 4 (17.4\%) &  ns$^{b}$ \\
IV & 25 (47.2\%) & 19 (36.5\%) & 8 (36.4\%) & 7 (30.4\%) &  ns$^{b}$\\
X & 18 (34.0\%) & 19 (36.5\%) & 11 (50.0\%) & 12 (52.2\%) &  ns$^{b}$\\
unknown & 1 (1.9\%) & 0 (0.0\%) & 0 (0.0\%) & 0 (0.0\%) &  ns$^{b}$\\
\multicolumn{1}{>{\footnotesize}l}{\textbf{T stage}} & \multicolumn{1}{>{\footnotesize}l}{} & \multicolumn{1}{>{\footnotesize}l}{} & \multicolumn{1}{>{\footnotesize}l}{} & \multicolumn{1}{>{\footnotesize}l}{} \\
T0 & 2 (3.8\%) & 0 (0.0\%) & 0 (0.0\%) & 0 (0.0\%) &  ns$^{b}$\\
T1 & 8 (15.1\%) & 8 (15.4\%) & 4 (18.2\%) & 2 (8.7\%) &  ns$^{b}$\\
T2 & 6 (11.3\%) & 10 (19.2\%) & 6 (27.3\%) & 6 (26.1\%) &  ns$^{b}$\\
T3 & 11 (20.8\%) & 13 (25.0\%) & 5 (22.7\%) & 10 (43.5\%)&  ns$^{b}$ \\
T4 & 23 (43.4\%) & 21 (40.4\%) & 6 (27.3\%) & 5 (21.7\%)&  ns$^{b}$ \\
TX & 2 (3.8\%) & 0 (0.0\%) & 1 (4.5\%) & 0 (0.0\%) \\
unknown & 1 (1.9\%) & 0 (0.0\%) & 0 (0.0\%) & 0 (0.0\%) &  ns$^{b}$\\
\multicolumn{1}{>{\footnotesize}l}{\textbf{N stage}} &  & \multicolumn{1}{>{\footnotesize}l}{} &  & \multicolumn{1}{>{\footnotesize}l}{} \\
N0 & 11 (20.8\%) & 20 (38.5\%) & 2 (9.1\%) & 10 (43.5\%) &  ns$^{b}$\\
N1 & 7 (13.2\%) & 9 (17.3\%) & 5 (22.7\%) & 3 (13.0\%) &  ns$^{b}$\\
N2 & 28 (52.8\%) & 18 (34.6\%) & 13 (59.1\%) & 9 (39.1\%) &  ns$^{b}$\\
N3 & 6 (11.3\%) & 5 (9.6\%) & 2 (9.1\%) & 1 (4.3\%) &  ns$^{b}$\\
NX & 0 (0.0\%) & 0 (0.0\%) & 0 (0.0\%) & 0 (0.0\%) &  ns$^{b}$\\
unknown & 1 (1.9\%) & 0 (0.0\%) & 0 (0.0\%) & 0 (0.0\%) &  ns$^{b}$\\
\multicolumn{1}{>{\footnotesize}l}{\textbf{M stage}} &  & \multicolumn{1}{>{\footnotesize}l}{} &  & \multicolumn{1}{>{\footnotesize}l}{} \\
M0 & 46 (86.8\%) & 50 (96.2\%) & 20 (90.9\%) & 22 (95.7\%) &  ns$^{b}$\\
M1 & 4 (7.5\%) & 2 (3.8\%) & 1 (4.5\%) & 0 (0.0\%) &  ns$^{b}$\\
MX & 2 (3.8\%) & 0 (0.0\%) & 1 (4.5\%) & 0 (0.0\%) &  ns$^{b}$\\
unknown & 1 (1.9\%) & 0 (0.0\%) & 0 (0.0\%) & 1 (4.3\%) &  ns$^{b}$\\ \hline
\hline
\end{tabular}
\captionsetup{justification=justified}
\captionsetup{justification=justified,font=footnotesize}
\vspace{5mm}
\end{table*}

\textbf{Fraction 5-specific model:} To build the fraction 5-ML, 102 patients with replanned fractions at fraction 5 and further were used as training\&validation set (52 non-replanned and 50 replanned). The best model for fraction 5 resulted from using the RF-RFE FS technique with 32 features. Figure 4 shows the ROC curves of the training/validation results and the final evaluation in the hold-out test set of 40 patients (23 non-replanned and 17 replanned). The selected features and hypermetrics are presented in Tables C and F. The results from the testing corresponded to an AUC of 0.82 and an accuracy of 82\%. In the confusion matrix, in 13 out of the 17 true positive cases, the actual request replan fraction was at fraction 5 or further. In the false negative cases, patients had requested replans at fractions 11, 16, 17, and 20; the reasons for replanning were changes in skin separation and weight loss. The resulting probabilities of not being replan were 51\%, 62\%, 78\% and 60\%, respectively.

\textbf{Fraction 10-specific model:} The best model for fraction 10 resulted from using the RF-RFE FS technique with 64 features, using  92 patients with replanned fractions at fraction 5, and further were used as training\&validation set (52 non-replanned and 40 replanned). The selected features and hypermetrics are presented in Tables D and F. The results from the testing with 36 hold-out patients (23 non-replanned and 13 replanned) corresponded to an AUC of 0.70 and an accuracy of 75\%. In the confusion matrix, in 7 out of the 13 true positive cases, the actual request replan fraction was at fraction 10 or further. In the false negative cases, patients had requested replan fractions at fractions 11, 13, 15, 16, 17, and 20. The reasons for replanning were PTV in air, swelling, and weight loss. The resulting probabilities of not being replan were 70\%, 52\%, 52\%, 58\%, 70\%, and 73\%, respectively. However, the patients of fractions 13 and 15 were successfully predicted previously with the fraction 5 model.

\textbf{Fraction 15-specific model:} To build the fraction 15-ML, 75 patients with replanned fractions at fraction 15 and further were used as a training\&validation set (22 replanned and 52 non-replanned). The best model for fraction 15 resulted from using the RF-RFE FS technique with 64 features. The selected features and hypermetrics are presented in Tables E and F. The results from the testing with 32 patients (23 non-replanned and 9 replanned) corresponded to an AUC of 0.79 and an accuracy of 84\%. In the confusion matrix, in 6 out of the 9 true positive cases, the actual request replan fraction was at fraction 10 or further. In the false negative cases, patients had requested replan fractions at fractions 16, 17, and 20. The resulting probabilities of not being replan were 84\%, 97\%, and 71\%. These patients were the same as the fraction 5-ML model was not able to classify. 

\subsection{ML specifications}
\vspace{5mm}

\begin{table}[H]
\captionv{12}{Short title - can be blank}{
\justifying{Type of selected geometrical metrics by the RF-RFE FS technique for the fraction 5-specific best model.
\label{Table_selection}}}
\vspace{-5mm}
\footnotesize
\centering
\begin{tabular}{ll}
\hline
\hline
 & \textbf{Related selected features for the best fraction 5 ML model} \\ \hline
\textbf{Clinical variables} & -   \\ \hline
\textbf{Values} & \makecell*[l]{$D_{Body}$, $CD_{Body}$, $CD_{Neck}$,$VI_{PTV}:LV_{Body}$,$VO_{PTV}:LV_{Body}$,\\$V^{air}_{Body-to-mask}$,$A^{2D}_{avg}$,$l_{y}$}\\ \hline
\textbf{Delta values} & $CD_{Body}$, $x^{PTV}_{avg}$,$x^{PTV}_{med}$,$\bar{B}_{mask}$,$M_{min}$,$A_{sub}$,$R^{sub}_{avg}$\\ \hline
\textbf{Slope values} &  $D_{Body}$,$VO_{PTV}:LV_{Body}$,$M_{min}$,$CD_{Neck}$,$\tilde{D}_{Neck}$,$R^{sub}_{min}$\\ \hline \hline
\end{tabular}
\end{table}
\vspace{5mm}

\begin{table}[H]
\captionv{12}{Short title - can be blank}{
\justifying{Type of selected geometrical metrics by the RF-RFE FS technique for the fraction 10-specific best model
\label{Table_selection2}}}
\vspace{-5mm}
\footnotesize
\centering
\begin{tabular}{ll}
\hline
\hline
 & \textbf{Related selected features for the fraction 10 ML model} \\ \hline
\textbf{Clinical variables} & - \\ \hline
\textbf{Values} & \makecell*[l]{$x^{PTV}_{max}$,$x^{PTV}_{std}$,$VI_{PTV}$,$VI_{PTV}:{LV_{Body}}$,$VO_{PTV}:LV_{Body}$,$CD_{Neck}$,\\$\tilde{D}_{Neck}$,$\bar{B}_{mask}$,$A_{sub}$}\\ \hline
\textbf{Delta values} & \makecell*[l]{$CD_{Body}$,$D_{Body}$,$VI_{PTV}:LV_{Body}$,$x^{PTV}_{min}$,$D_{Neck}$,$\bar{D}_{Neck}$,$\tilde{D}_{Neck}$,$M_{avg}$,\\$A^{2D}_{avg}$,$A_{sub}$,$\varphi^{2D}_{R^{sub}}$} \\ \hline
\textbf{\makecell*[l]{Slope values\\\hspace{1mm}}}&  \makecell*[l]{$D_{Body}$,$CD_{Body}$,$V_{Body}$$x^{PTV}_{min}$,$x^{PTV}_{avg}$,$x^{PTV}_{med}$,$VO_{PTV}$,$VI_{PTV}:LV_{Body}$,\\$VO_{PTV}:LI_{Body}$,$D_{Neck}$,$\tilde{D}_{Neck}$,$V_{Neck}$,$M_{min}$,$A^{2D}_{avg}$,$A_{sub}$} \\ \hline \hline
\end{tabular}
\end{table}

\vspace{10mm}

\begin{table}[H]
\captionv{12}{Short title - can be blank}{
\justifying{Type of selected geometrical metrics by the RF-RFE FS technique for the fraction 15-specific best model.
\label{Table_selection3}}}
\vspace{-5mm}
\footnotesize
\centering
\begin{tabular}{ll}
\hline
\hline
 & \textbf{Related selected features for the fraction 15 ML model} \\ \hline
\textbf{Clinical variables} & - \\ \hline
\textbf{Values} & $x^{PTV}_{min}$,$x^{PTV}_{max}$,$x^{PTV}_{avg}$,$VO_{PTV}:LI_{Body}$,$max\{B_{mask}\}$,$R^{3D}_{max}$,$\varphi^{2D}_{R^{sub}}$\\ \hline
\textbf{Delta values} & $x^{PTV}_{min}$,$\varphi^{2D}_{R^{sub}}$ \\ \hline
\makecell*[l]{\textbf{Slope values}\\\hspace{1mm}}& \makecell*[l]{$x^{PTV}_{min}$,$x^{PTV}_{med}$,$x^{PTV}_{avg}$,$VI_{PTV}:LI_{Body}$,$M_{avg}$,$M_{std}$,$\bar{B}_{mask}$,$V_{Neck}$,\\$SA_{Neck}$,$R^{3D}_{avg}$,$A^{2D}_{avg}$,$A_{sub}$,$R^{sub}_{min}$}\\ \hline
\hline
\end{tabular}
\captionsetup{justification=justified}
\label{Table_selection3}
\end{table}
\vspace{10mm}

\begin{table}[H]
\captionv{12}{Short title - can be blank}{
\justifying{Hyperparameters selected during the hyperparameter-tuning process using the GridSearchCV procedure. The candidate parameters started from the default values provided by \texttt{scikit-learn}.
\label{Hyper}}}
\vspace{-5mm}
\footnotesize
\centering
\begin{tabular}{cccc}
\hline
\hline
\multicolumn{1}{l}{\textbf{Classifier}} & \multicolumn{1}{l}{\textbf{Treatment fraction}} & \textbf{Hyperparameters} & \textbf{Results} \\ \hline
\multirow{2}{*}{ET} & \multirow{2}{*}{5} & criterion & \textit{entropy} \\
 &  & maximum number features & \textit{sqrt} \\
\multirow{4}{*}{ET} & \multirow{4}{*}{10} & number of estimators & 100 \\
 &  &criterion & \textit{gini} \\
 &  & maximum number features & \textit{sqrt} \\
 &  & number of estimators & 100 \\
\multirow{2}{*}{SVM} & \multirow{2}{*}{15} & kernel & \textit{sigmoid}\\ & &C (regularization parameter) & 1 \\
 &  & gamma (kernel coefficient) & auto \\ \hline \hline
\end{tabular}

\end{table}
\end{document}